\documentclass[12pt,aps,prc,amssymb,floatfix]{revtex4}
\usepackage{graphicx}
\usepackage{epsfig}    

\def\lsim{\mathrel{\rlap{
\lower4pt\hbox{\hskip-3pt$\sim$}}
    \raise1pt\hbox{$<$}}}     
\def\gsim{\mathrel{\rlap{
\lower4pt\hbox{\hskip-3pt$\sim$}}
    \raise1pt\hbox{$>$}}}     

\begin{document}

\begin{center}
{\large {\bf QCD Matter: A Search for a Mixed Quark-Hadron Phase}}

 \vspace*{5mm}

A.N.~Sissakian, A.S.~Sorin, and V.D.~Toneev \\
\vspace*{3mm}

{\em Bogoliubov Laboratory of Theoretical Physics,}\\
{\em Joint Institute for Nuclear Research (JINR),}\\
{\em 141980, Dubna, Moscow Region, Russia}\\
{\tt sisakian@jinr.ru ~~ sorin@theor.jinr.ru ~~ toneev@theor.jinr.ru}

\end{center}

\vspace*{5mm}

{\small { \centerline{\bf Abstract} Physics aspects of a JINR
project to reach the planned 5A GeV energy for the Au and U beams
and to increase the bombarding energy up to 10A GeV are discussed.
The project aims to search for a possible formation of a strongly
interacting mixed quark-hadron phase. The relevant problems are
exemplified. A need for scanning heavy-ion interactions in
bombarding energy, collision
centrality and isospin asymmetry is emphasized. }}\\[5mm]

\section{Introduction}

Over the last 25 years a lot of efforts have been made to search
for new states of strongly interacting matter under extreme
conditions of high temperature and/or baryon density, as predicted
by Quantum Chromodynamics (QCD). These states are relevant to
understanding the evolution of the early Universe after Big Bang,
the formation of neutron stars, and the physics of heavy-ion
collisions. The latter is of great importance since it opens a way
to reproduce these extreme conditions in the Earth laboratory.
This explains a permanent trend of leading world research centers
to construct new heavy ion accelerators for even higher colliding
energy.

Looking  at the list of heavy-ion accelerators one can see that
after the pioneering experiments at the Dubna Synchrophasotron,
heavy-ion physics developed  successfully at Bevalac (Berkley)
with the bombarding energy to $E_{lab} \sim 2$A GeV, AGS
(Brookhaven) $E_{lab} \sim 11$A GeV, and SPS (CERN) $E_{lab} \sim
160$A GeV. The first two machines are closed now. The nuclear
physics programs at SPS as well as at SIS (GSI, Darmstadt,
$E_{lab} \sim 1$A GeV) are practically completed. The new
relativistic heavy-ion
 collider (RHIC, Brookhaven) is intensively working in the
ultrarelativistic energy range $\sqrt{s_{NN}}\sim 200$ GeV to
search for signals of the quark-gluon plasma formation. In this
respect, many hopes are related to the Large Hadron Collider (LHC,
CERN) which will start to operate in  the TeV region in two-three
years. The low-energy scanning program at SPS (NA49 Collaboration)
revealed an interesting structure in the energy dependence of some
observables at $E_{lab} \sim 20-30$A GeV which can be associated
with the exit of an excited system into a deconfinement state.
This fact essentially stimulates a new large project FAIR GSI
(Darmstadt) for studying compressed baryonic matter in a large
energy range of $E_{lab} =10-30$A GeV which should come into
operation after 2015 year~\cite{GSI300}. These problems are so
attractive that the RHIC scientific society discusses a
possibility to decrease the collider energy till the FAIR
range~\cite{RHIC06}.

On the other hand, in JINR there is a modern superconducting
accelerator, Nuclotron, which has not realized its planned
parameters yet. The Veksler and Baldin Laboratory of High Energy
has certain experimental facilities and large experience in
working with heavy ions. This study may actively be supported by
theoretical investigations of the Bogoliubov Laboratory of
Theoretical Physics. In~\cite{SSSTZ06} a program  was proposed for
investigating the dense strongly interacting QCD matter, formed in
relativistic heavy ion collisions, based on acceleration of heavy
ions like Au at the Nuclotron up to the maximal planned energy
$E_{lab}=5$A GeV.  In view of new opened opportunities of the
Nuclotron update  to increase the bombarding energy up to 10A GeV
and to get both Au and U ions with relativistic energies, the
relevant physics problems are discussed in this paper.

\section{Phase diagrams}
\begin{figure}[h]
  \includegraphics[width=7.3cm]{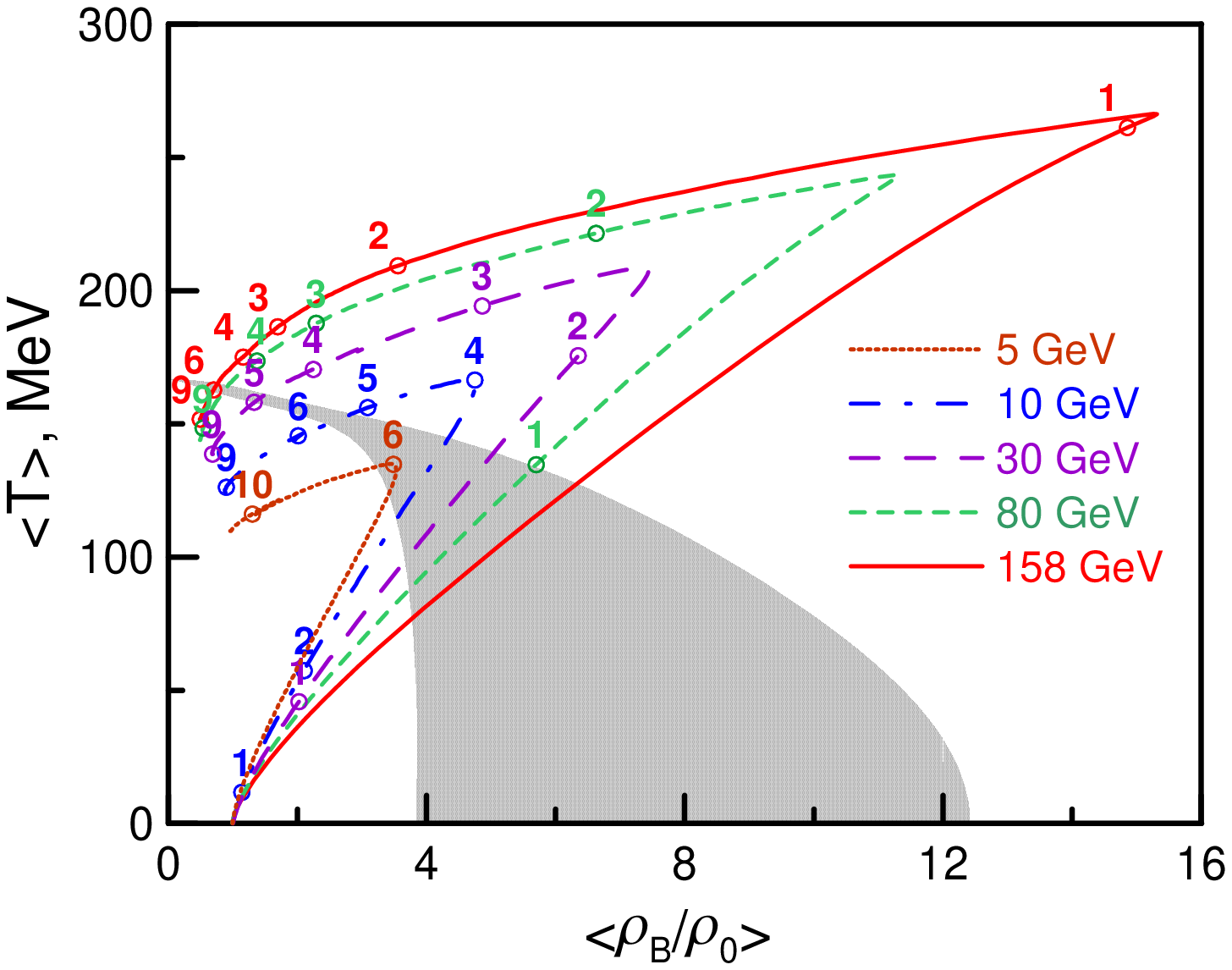}
  \includegraphics[width=7.3cm]{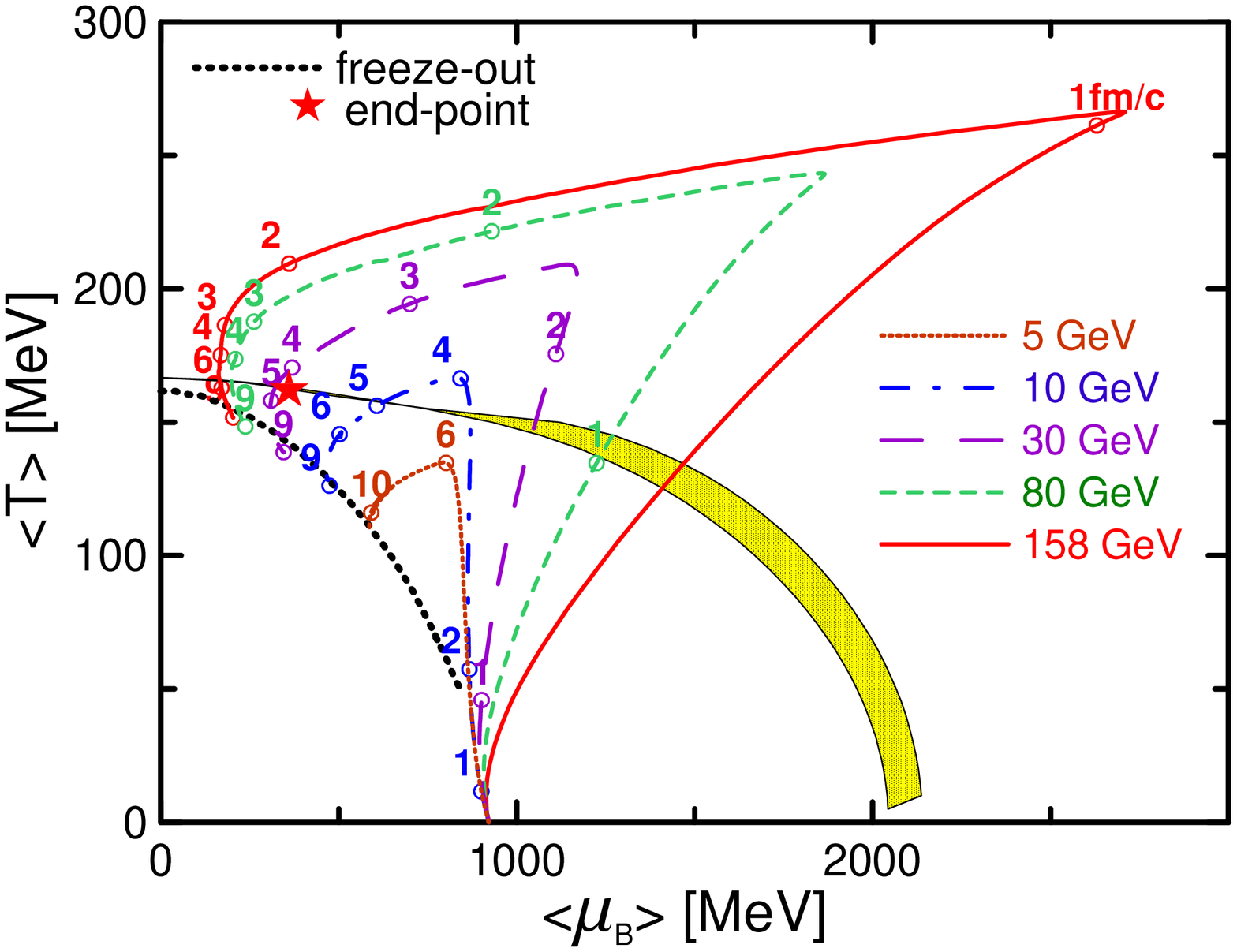}
  \caption{Dynamical trajectories for central ($b=2$ fm) Au+Au collisions
  in $T-\rho_B$ (left panel) and  $T-\mu_B$ (right panel) plane for
  various bombarding energies calculated within  the relativistic 3-fluid
  hydrodynamics calculated with hadronic EoS~\cite{IRT05}. Numbers
  near the trajectories are the evolution time moment.  Phase boundaries are
  estimated in a two-phase bag model~\cite{KSTR06}. In the right
  panel the critical end-point calculated in the lattice QCD~\cite{Fodor01}
   is marked by the star
  and the shaded region corresponds to uncertainties of the bag model.
  }
  \label{fig2-3j}
\end{figure}

A convenient way to present a variety of possible states of
strongly interacting matter is a phase diagram in terms of
temperature $T$ and baryon chemical potential $\mu_B$ (or baryon
density $\rho_B$), as presented in Fig.\ref{fig2-3j}. This
picture shows in which region of the diagram the given phase is
realized and which colliding energies are needed to populate this
region.

As is seen, a system, formed in a high energy collision, is fast
heated and compressed and then starts to expand slowly reaching
the freeze-out point which defines observable hadron quantities.
At the Nuclotron energy $E_{lab}=5$A GeV the system "looks" into
the mixed phase  for a short time (the left part of
Fig.\ref{fig2-3j}), however, uncertainties of these calculations
are still large. To get agreement with the lattice data for finite
$T$ and $\mu_B$, masses of $u,d$ quarks should be rather heavy,
and the phase boundary is shifted towards higher
$\mu_B$~\cite{KSTR06} (the right part in Fig.\ref{fig2-3j}). On
the other hand, one should exercise caution because these
dynamical trajectories were calculated for a pure hadronic gas
equation of state, and the presence of a phase transition may
noticeably change them. In addition, near the phase transition the
strongly interacting QCD system behaves like a liquid rather than
a gas, as was clarified recently at small $\mu_B$ from both
quark~\cite{ZS} and hadronic~\cite{V04} side. As to high $\mu_B$
values, it is a completely open question. One should  also note
that as follows from lattice QCD calculations for $\mu_B \approx
0$ the deconfinement temperature practically coincides with the
transition temperature for chiral symmetry  restoration while for
baryon-rich matter it is still an open question.

\begin{figure}[h]
  \includegraphics[width=9cm]{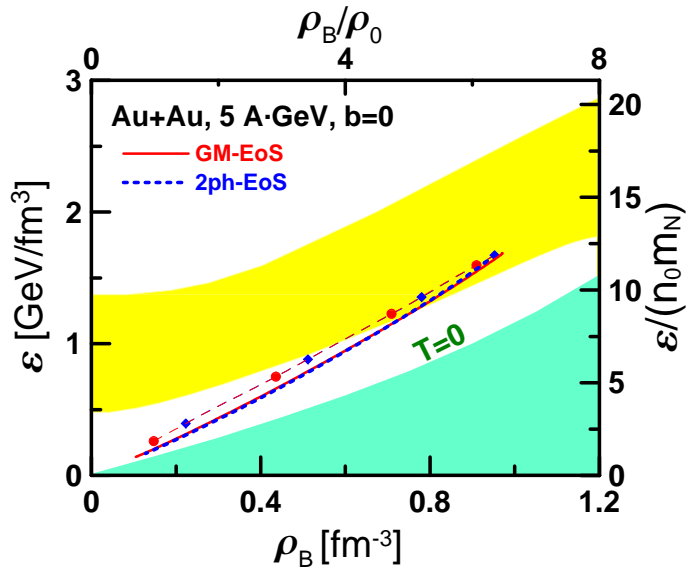}
  \caption{Dynamical trajectory in the $\varepsilon -\rho_B$-plane for
  central Au+Au collisions calculated with
  two equations of state: pure hadronic (solid line) and with first
  order phase transition (dashed). Spatial averaging is done over
  the cube with the 4 fm sides and Lorentz contracted in the
  longitudinal direction~\cite{I05}. The shaded regions
  correspond to the mixed phase (upper one) and the non-reachable
  domain with the boundary condition $T=0$, respectively.
  }
  \label{fign_eps}
\end{figure}

In Fig.\ref{fign_eps}, dynamical trajectories in the
$\varepsilon-\rho_B$ plane for the top project Nuclotron energy
are given for two equations of state, without and with the
first-order phase transition~\cite{IRT05,I05}. The shaded band
corresponds to the mixed phase of more restrictive EoS with heavy
masses of $u,d$ quarks, as shown in the right panel of
Fig.\ref{fig2-3j}. Nevertheless, both the trajectories, being
close to each other, spend some time in the mixed phase. The main
difference between the results presented in Fig.\ref{fig2-3j} and
Fig.\ref{fign_eps} comes from the different averaging procedures
used: in the first case, thermodynamic quantities were averaged
over the whole volume of the interacting system, while in the
second case, it was carried out only over a cube of 4 fm sides
placed at the origin being Lorentz contracted along the colliding
axis. Thus, for central collisions at the 5A GeV Nuclotron energy
even if an average state of the whole strongly interacting system
does not approach the mixed phase, an essential part of the system
volume will spend a certain time in this mixed phase. An
experimental consequence  is that an expected observable signal of
reaching the mixed phase should be rather weak. Note that for
$E_{lab}=$10A GeV these conditions for a phase transition are
fulfilled appreciably better.

\begin{figure}[h]
  \includegraphics[width=8.5cm]{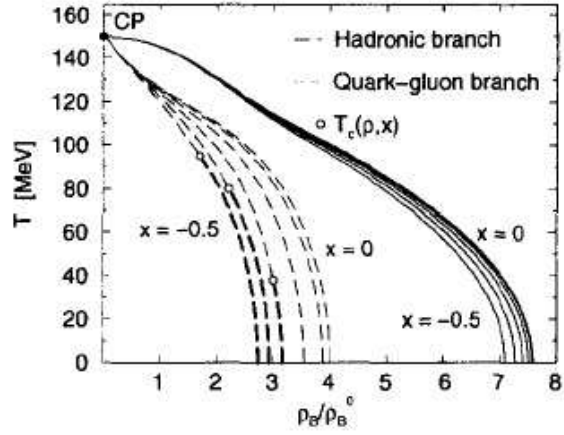}
 \caption{Projections
  of the boundary surface  on the $(\rho_B,T)$ plane
  at different $x$. The isospin ratios  are $x=0,
  -0.1, -0.2, -0.3, -0.4, -0.5$, starting from the right.
  The boundaries of the mixed phase with the quark and hadron
  phase are plotted as
  solid and dashed lines, respectively~\cite{M97}.}
  \label{rhoB_T}
\end{figure}

One should stress that the presented above dynamical trajectories
and boundaries for the first-order phase transition were estimated
for a system conserving a single charge, namely the baryonic
charge. However, the behavior of this system near the phase
transition and particularly within the mixed phase will be
qualitatively different if conservation of more than one charge is
taken into account~\cite{LL69,G92}. So we turn to consideration of
effects of additional conservation of the electric charge, or
isospin of the system.

We shall characterize the charge asymmetry by the
electric-to-baryonic charge ratio $Z/A=\rho_Q/\rho_B$ or by the
isospin ratio $x=(\rho_Q-\rho_B)/2\rho_B$ which are related as
$x=Z/A-0.5$. For Au+Au collisions we have $Z/A=0.4$ or $x=-0.1$.

An essential difference in the first-order phase transition for
systems with one and two conserved charges stems from the fact
that the phase boundary, which is a line for a single charge
conservation, is getting   a two-dimensional surface. In the
latter case the conserved charges can now be shared by two phases
in equilibrium in different concentration in each phase but
consistent with the global charge conservation. If this is
energetically favored by the internal forces and Fermi energies,
then these degrees of freedom will be exploited by the system and
will influence thermodynamic quantities.

The boundary surface, so called binodal, may be parameterized in a
different way, say as $\{T,\mu_B,\mu_Q\}$, $\{T,\mu_B, x\}$,
$\{T,\rho_B, x\}$ or $\{p,\rho_B, x\}$. In Fig.\ref{rhoB_T}, some
$\{T,\rho_B, x\}$ projections of a hadron-quark phase transition
are shown~\cite{M97}.

It is seen that, for example, in symmetric matter  at $T=0$ the
baryon density ranges along 3.5$\rho_0$ between the onset and
completion of the transition. For $x=0$ the hadron boundary is
close to that in the left panel of Fig.\ref{fig2-3j}. This mixed
phase domain becomes even larger in density for iso-asymmetric
systems. Another important observation is that the density at the
onset (i.e. hadronic side of the phase boundary at the transition
density $\rho_c$) decreases with increasing isospin asymmetry. If
one compares  points for $x=0$ and $x=-0.2$ at $T=0$, a decrease
is $\delta \rho_B \approx 0.5 \rho_0$.  But this effect is
practically absent for $T\gsim 120$ MeV.

\begin{figure}[h]
  \includegraphics[width=8.5cm]{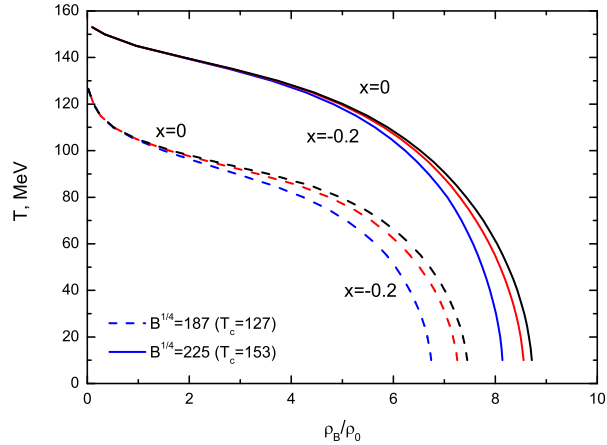}
 \caption{Hadron boundary of the mixed phase
  at different $x$~\cite{KTV06}. The isospin ratios  are $x=0,
  -0.1, -0.2$, starting from the right. Hadronic phase is
  described within the relativistic mean-field approach
  with density-dependent hadron masses and coupling constants~\cite{KV05}.
  The results are given for two values of the bag constant: $B^{1/4}=$187
  (dashed lines) and 225 MeV (solid lines).
  }
  \label{bnd}
\end{figure}

A more realistic description of the hadronic phase, which takes
into account the density dependence of hadron masses and coupling
constants~\cite{KV05}, was used for the results presented in
Fig.\ref{bnd}. The general trend of curves is quite similar to
that in Fig.\ref{rhoB_T} but now the values of transition
densities are higher though the same bag constant $B^{1/4}=$187
MeV was used in both the calculations~\cite{M97,KV05}. The main
difference comes from the very low mass used for in-medium nucleon
at the normal nuclear density and $T=0$, $M^*_N=0.6M_N$. In this
case the coefficient $C_4$ ahead the highest sigma-meson
interaction term $C_4\sigma^4$ turns out to be negative and,
therefore, such a system should be unstable~\cite{KV05}. If the
bag constant increases, the transition boundary moves to higher
values. Note that in both the cases the temperature $T(\mu_B=0)$
is lower than the value 170-180 MeV expected from the lattice
calculations. However, the lattice QCD data show that we deal with
the first-order phase transition only for the high baryon chemical
potential, above that for the critical end-point~\cite{Fodor01}
(see also the right panel in Fig.\ref{fig2-3j}). Thus, it is not
completely clear whether two-phase calculations should describe
the lattice value of $T(\mu_B=0)$. Appropriate lattice QCD data
are available only for small $\mu_B$ and there are no data taking
into consideration the conservation of both baryon and electric
charges.

\begin{figure}[h]
 \includegraphics[width=6.5cm,angle=90]{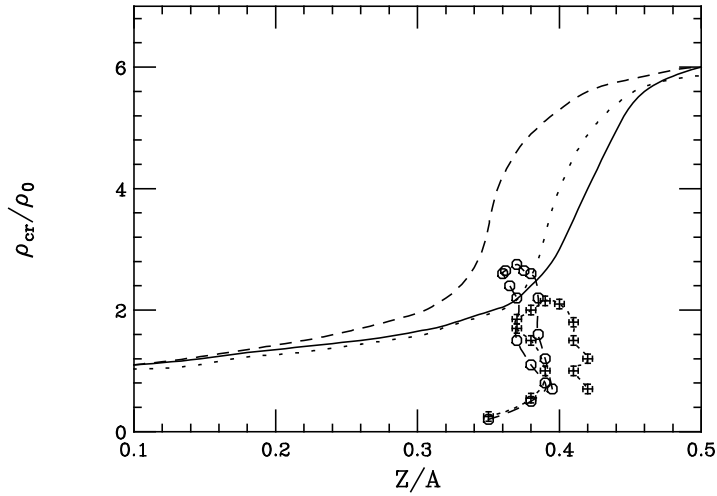}
  \caption{Variation of the transition density with the proton fraction
  at $T=0$ for various EoS parameterizations of the relativistic mean-field
  theory: Dotted, dashed and solid lines correspond to the GM3
  version~\cite{GM91}, to an additional inclusion of non-linearity
  in $\rho$ interaction~\cite{LGBCDiT02} and to an extra term
  describing the interaction of the isovector $\delta$-meson~\cite{DiTDGGL06},
  respectively. The points represent the dynamical trajectory in
  interaction zone during semi-central $^{132}Sn~+~^{132}Sn$ collisions
  at 1A GeV (circles) and at 300A MeV (crosses)~\cite{DiTDGGL06}. }
  \label{fig4j}
\end{figure}

Even noticeably stronger reduction of the transition density with
increasing neutron fraction is predicted by the Catania
group~\cite{DiTDGGL06} for  $T=0$. It was demonstrated that
$\rho_c$ depends appreciably on poorly known properties of EoS at
high baryon densities. This statement is illustrated in
Fig.\ref{fig4j}. Note quite low values of the transition densities
which originate from small values of the bag constant used,
$B^{1/4}=140-170$ MeV. This choice is argued by the following
Witten hypothesis~\cite{W84}: a state made of an approximately
equal number of $u, d, s$ quarks can have the energy per baryon
number smaller than that for Fe; therefore, the quark matter is
absolutely stable and purely quark stars may exist. This
hypothesis put strong constraints on quark model parameters: The
bag model parameter $B$ should be small~\cite{W84}. Today the
existence of pure quark stars is not excluded but is rather
considered as an exotic case. In this respect, the
(non)observation of a strong isospin dependence of the transition
density may be treated as a test of the Witten hypothesis.

The most striking feature of all results is a sharp decrease of
the transition density $\rho_c$ which takes place in the range
$Z/A\sim$ 0.3$\div$0.45, though the size of the reduction effect
and its position on the $Z/A$ axis, as is seen in Fig.\ref{fig4j},
are strongly model dependent. Nevertheless, application of
neutron-rich heavy ions seems to be very perspective to study the
mixed QCD matter at energies even lower than the project Nuclotron
energy 5A GeV. Dynamical trajectories presented in Fig.\ref{fig4j}
show that the phase boundary may be reached at energies as low as
1A GeV.
\begin{figure}
\begin{center}
\includegraphics[angle=-90,width=14.cm]{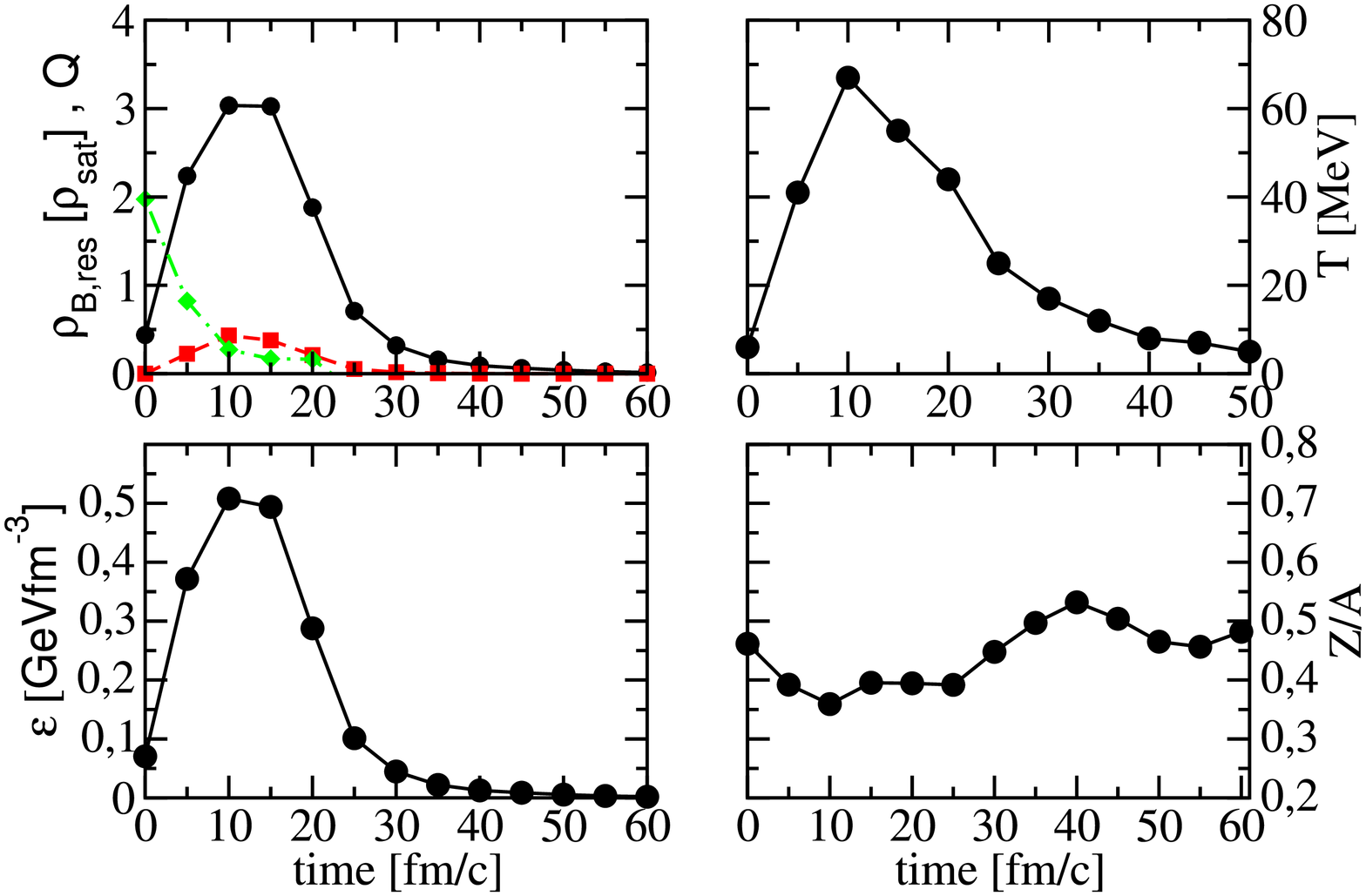}
\caption{\label{figUU} Time evolution of thermodynamic quantities
inside a cubic cell of 2.5 fm wide, located in the center of mass
of the system, is shown for semi-central $^{238}U-^{238}U$
collisions at $E_{lab}=$1A GeV with $b=$7 fm. Baryon density,
temperature, energy density and proton fraction are presented.
Different curves in the upper-left panel are: {\it black dots} --
the baryon density in $\rho_0$ units; {\it grey dots} -- the
quadruple momentum in momentum space; {\it squares} -- the
resonance density~\cite{DiTDGGL06}.} 
\end{center}
\end{figure}

 As follows from Figs.\ref{rhoB_T}-\ref{fig4j}, the most
favorable temperatures for this reduction effect are in the range
$T\lsim$ 80 MeV. It is well known that in central collisions of
relativistic heavy ions the growth of the density $\rho_B
> 3 \rho_0$ with the energy increase is accompanied by the appropriate
rise in temperature $T\gsim 100$ MeV. Available stable nuclei
cover only a very narrow region in isospin asymmetry $Z/A\approx$
0.39$\div$0.40 exhausted by the $^{238}_{~92}U$ and
$^{197}_{~79}Au$ isotopes, respectively, what embarrasses checking
the boundary reduction effect. The use of the long-lived
$^{195}_{~79}Au$ isotope ($\tau_l\sim$150 days) allows one to move
towards neutron-poor side of the boundary till $Z/A=$0.41. In
addition, the construction of accelerators of intense beams of
neutron-rich heavy ions with $Z/A$ beyond this narrow region is an
extremely complex technological problem. \cite{DiTDGGL06}.
 Some possibility to get
lower temperatures and to extend the reached isospin asymmetry
region is opened by study of semi-central rather than central
collisions. This possibility is illustrated in Fig.\ref{figUU}

   It is noteworthy that after about $10$ fm/$c$ the quadruple
momentum is almost vanishing and a nice local equilibration is
achieved. At this beam energy the maximum density coincides with
reaching the thermalization. Then the system is quickly cooling
while expanding.  In Fig.\ref{figUU} one can see that rather
exotic nuclear matter is formed in a transient time of the order
of $10$ fm/$c$, having the baryon density around $3\rho_0$, the
temperature $50-60~MeV$, the energy density $500~MeV$~fm$^{-3}$ and
the proton fraction between $0.35$ and $0.40$. So the local
neutron excess well inside the estimated mixed phase region may be
even higher than that in colliding nuclei (note that for $^{238}U$
the isospin symmetry ratio is $Z/A=$0.387).
\begin{figure}[h]
 \includegraphics[width=15.5cm]{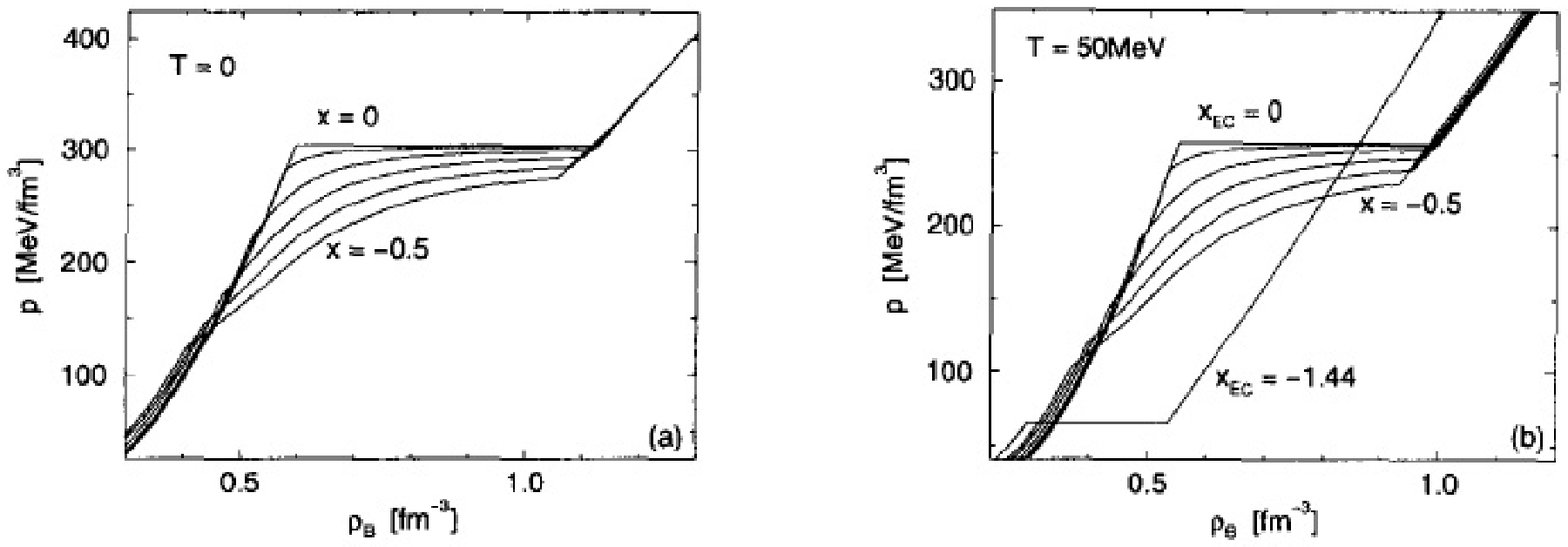}
  \caption{Isoterms for different values of $x$ at $T=0$ (left panel)
  and $T=50$ MeV (right panel). The isospin ratios  are $x=0, -0.1, -0.2,
  -0.3, -0.4, -0.5$ from top to bottom~\cite{M97}. }
  \label{p-rhoB}
\end{figure}

The $\{p,\rho_B, x\}$ projections of EoS corresponding to the
Maxwell construction are given in Fig.\ref{p-rhoB} for two
isotherms. In accordance with the result familiar from the
behavior of the systems with one conserved charge, for symmetric
matter ($x=0$) the pressure stays constant within the mixed phase.
In contrast, at $x<0$ the pressure changes during the transition
increasing with the baryon density. This change of the pressure
throughout the phase separation in asymmetric systems is an
indication of a smoother transition than in symmetric systems as
was first noted in the context of neutron star
calculations~\cite{G92}. Certainly, such behavior of pressure may
crucially affect the evolution of two colliding nuclei.

\begin{figure}[h]
 \includegraphics[width=8.5cm]{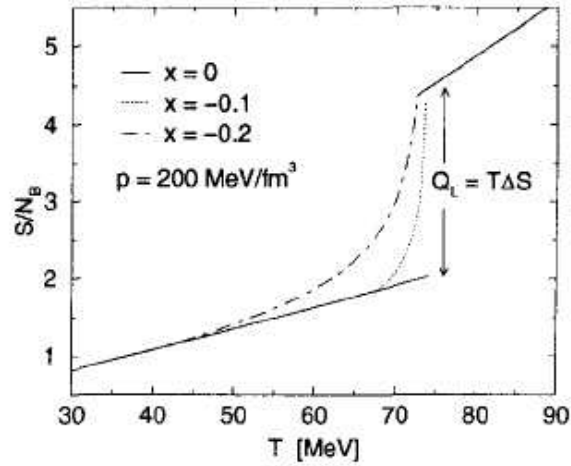}
  \caption{Entropy per baryon as a function of temperature at the
  constant pressure for various isospin ratios~\cite{M97}. For the
  case $x=0$ the latent heat $Q_L$ is shown.}
  \label{S_T}
\end{figure}
\begin{figure}[h]
 \includegraphics[width=7.5cm]{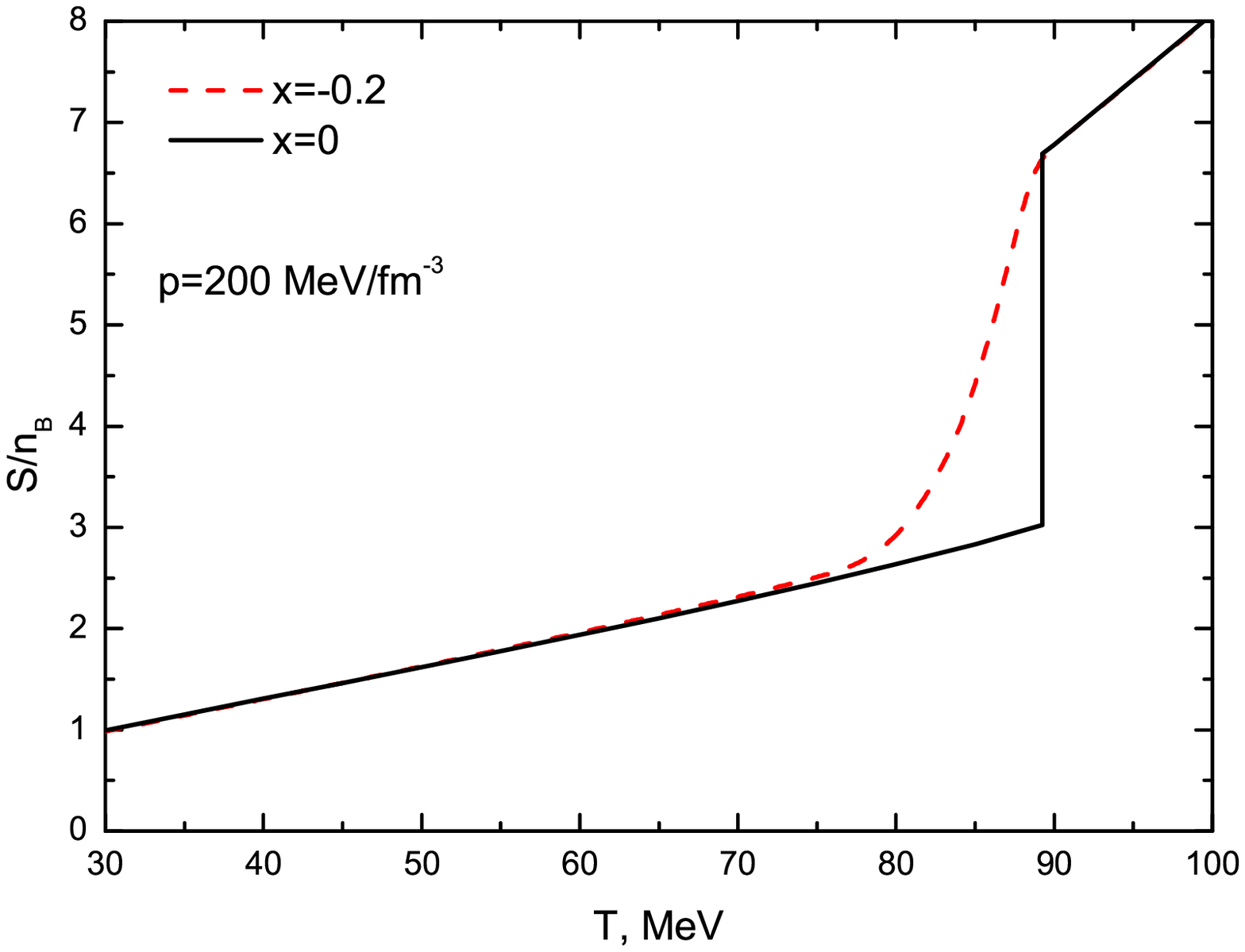}
 \includegraphics[width=7.5cm]{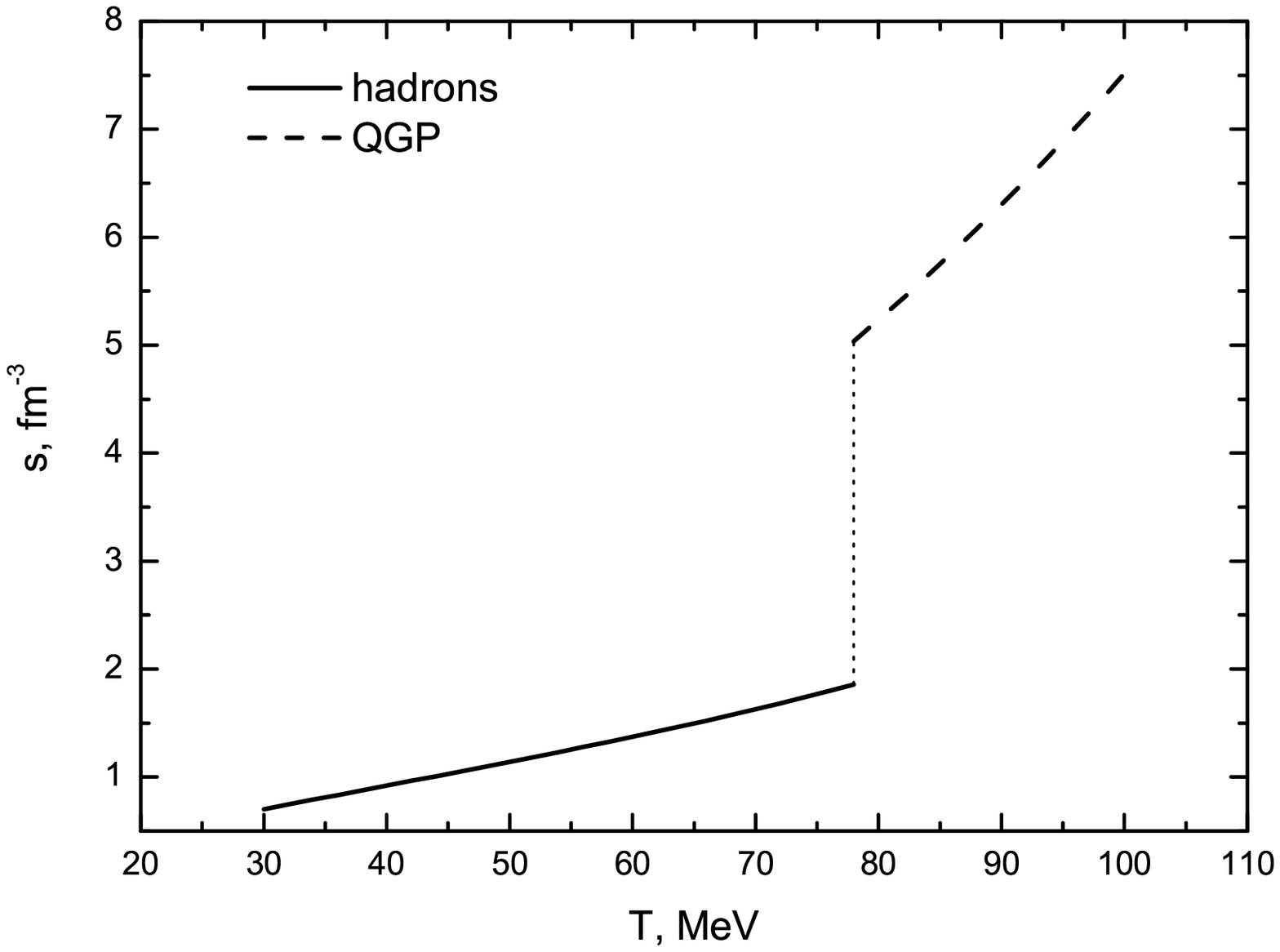}
 \caption{Entropy per baryon as a function of temperature at the
  constant pressure for various isospin ratios (left panel) and entropy
  as a function of temperature at the constant chemical potentials
  $\mu_B=$1300, $\mu_S=$300 and $\mu_Q=$-100 MeV (right panel). The
  relativistic mean-field model with the density-dependent
   interaction is used for the hadronic phase and the bag model describes the
   quark phase~\cite{KTV06}.}
  \label{Smu_T}
\end{figure}

\begin{figure}[h]
 \includegraphics[width=7.2cm]{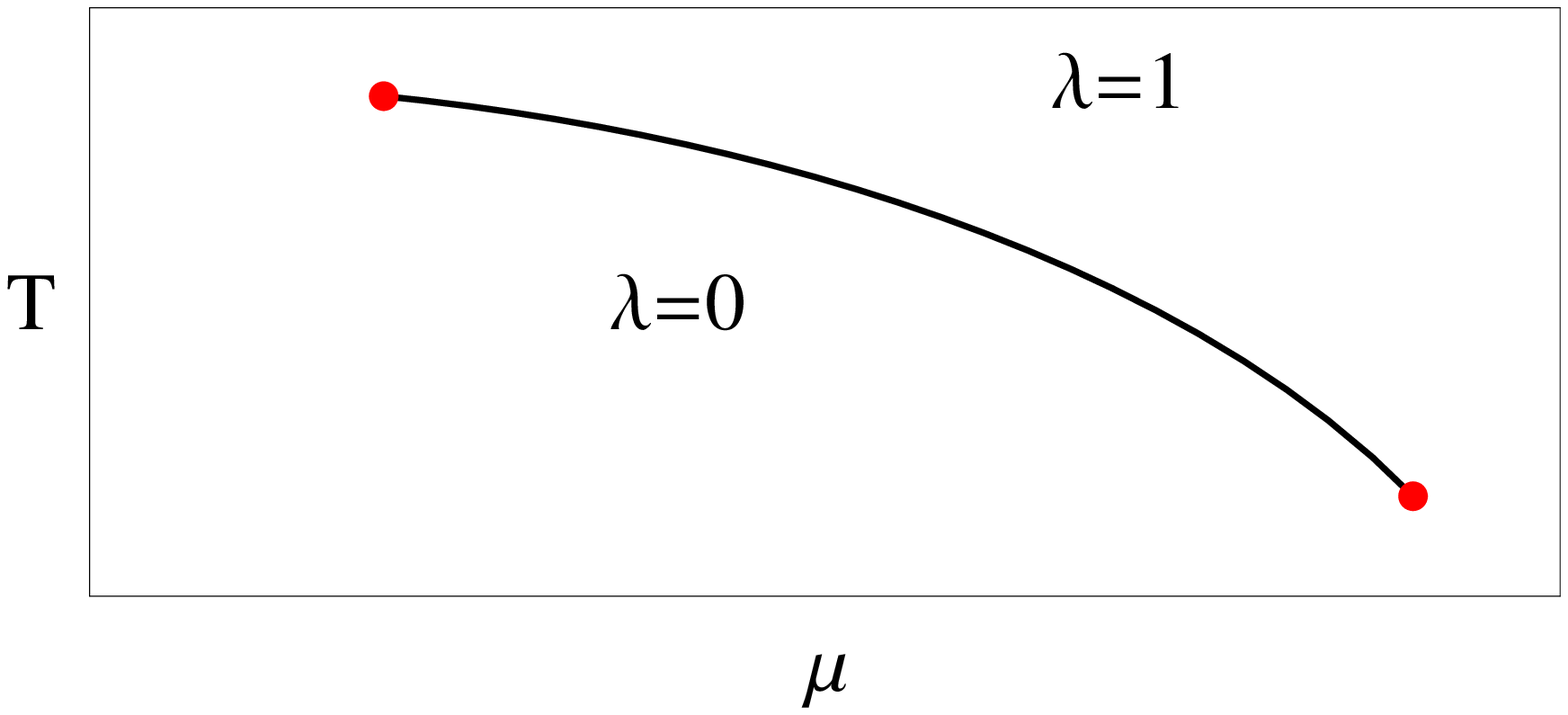}\hspace*{5mm}
 \includegraphics[width=7.2cm]{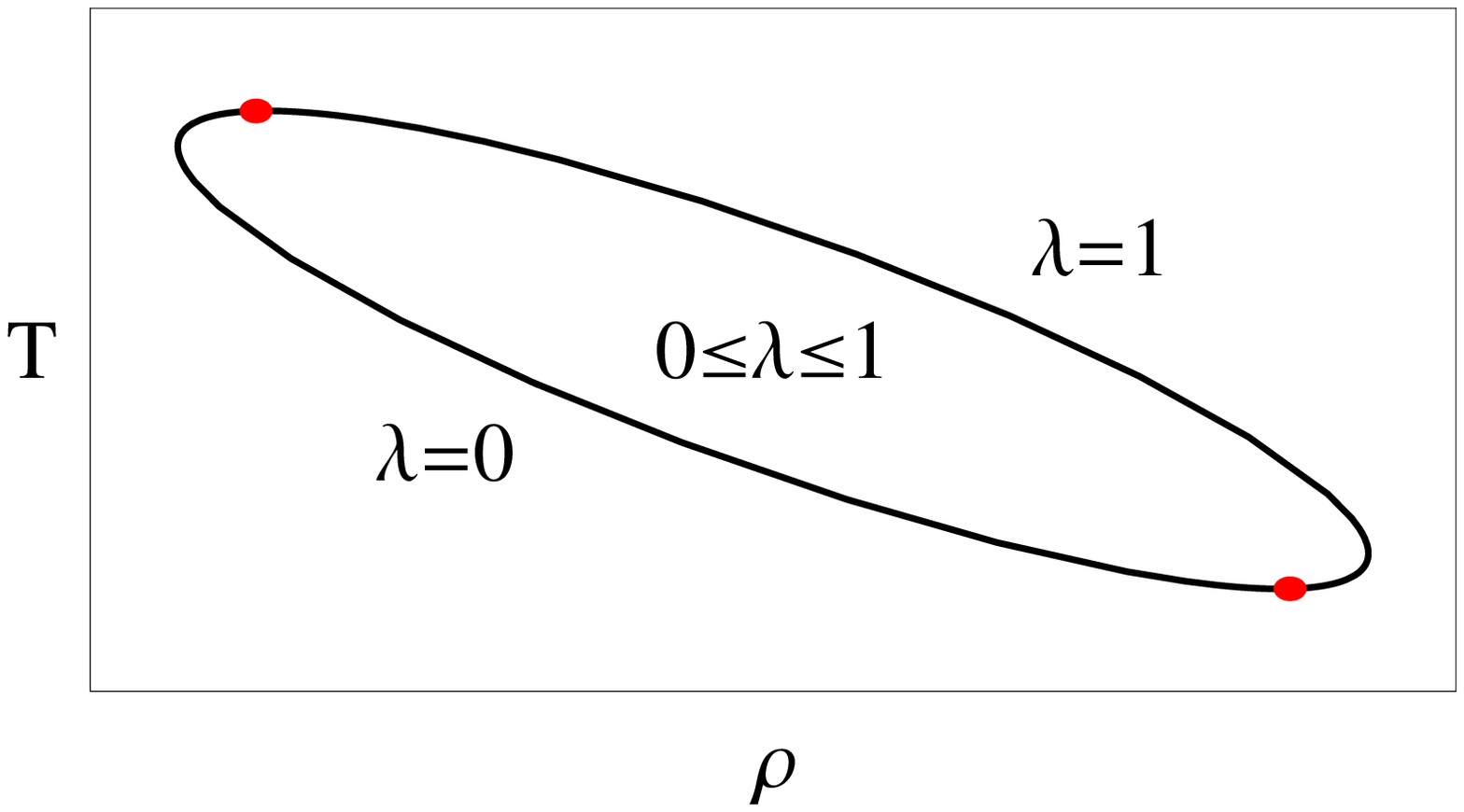}\\
\includegraphics[width=7.2cm]{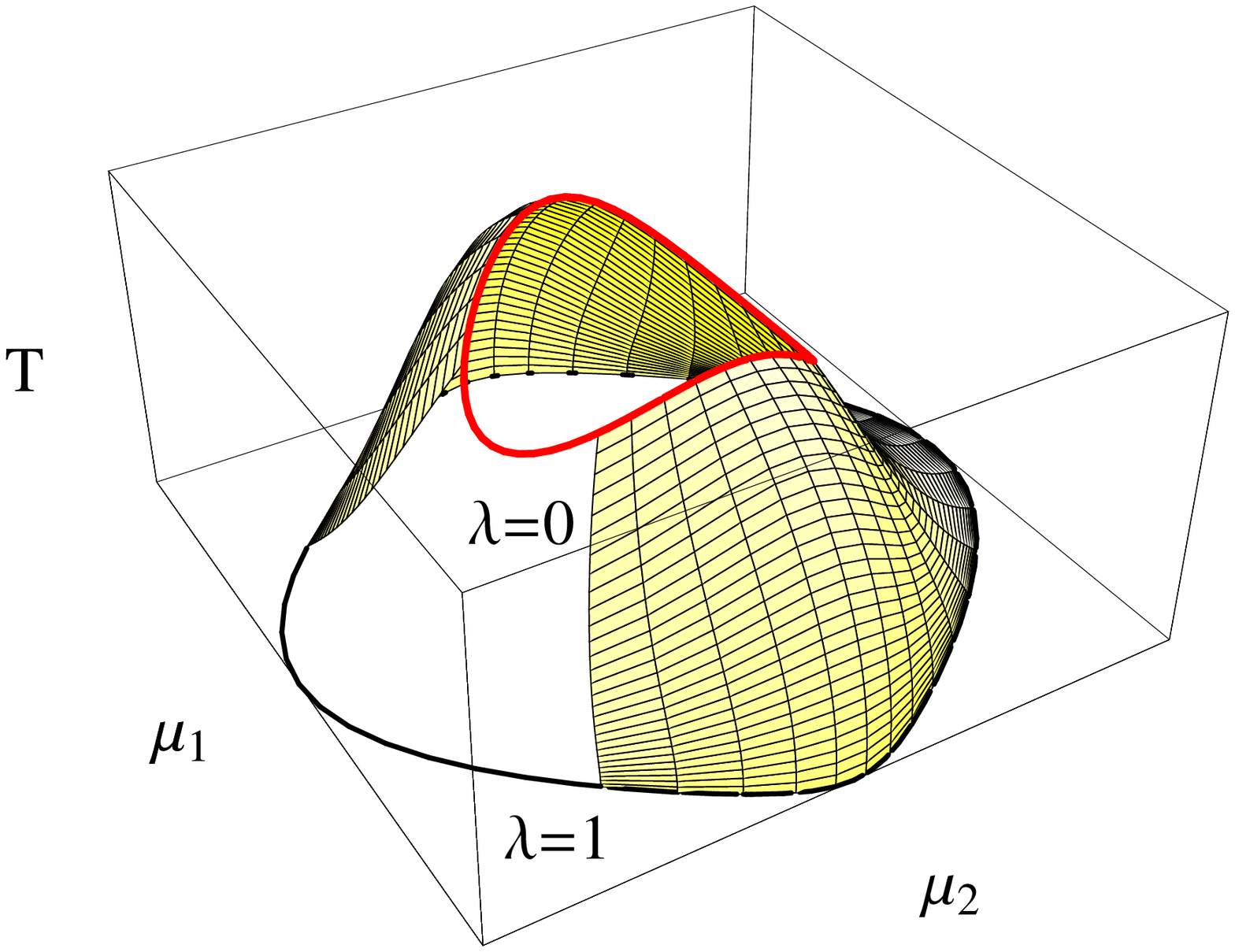}\hspace*{5mm}
 \includegraphics[width=7.2cm]{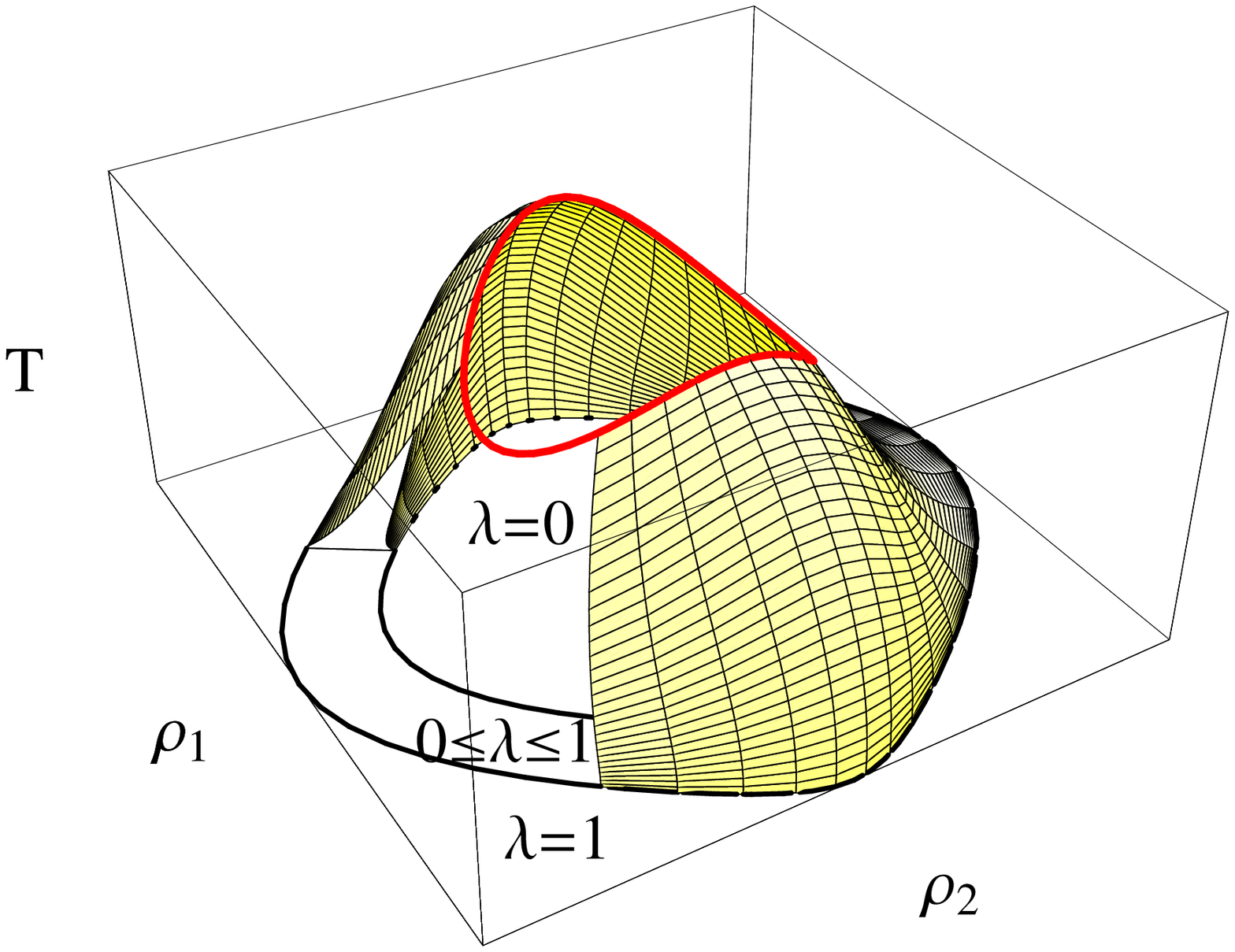}\\
\includegraphics[width=7.2cm]{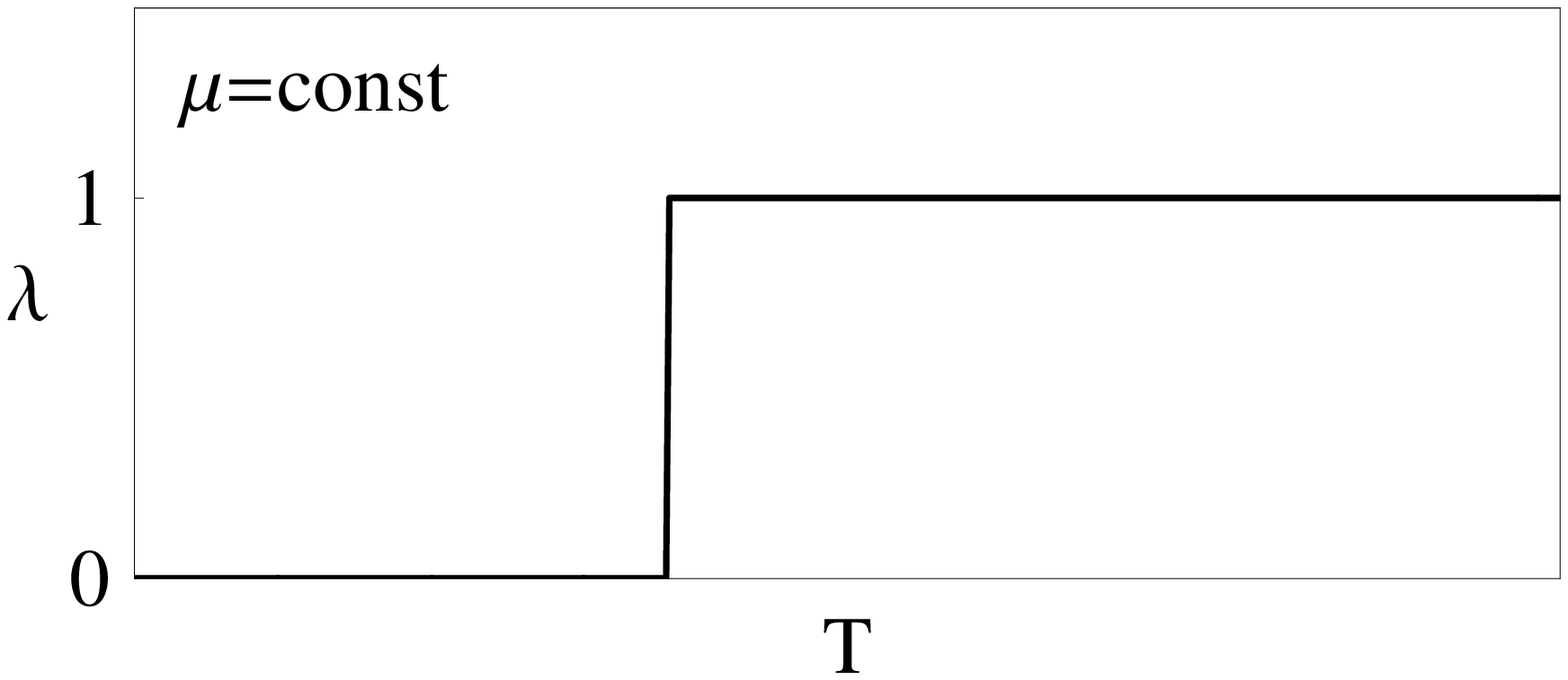}\hspace*{5mm}
 \includegraphics[width=7.52cm]{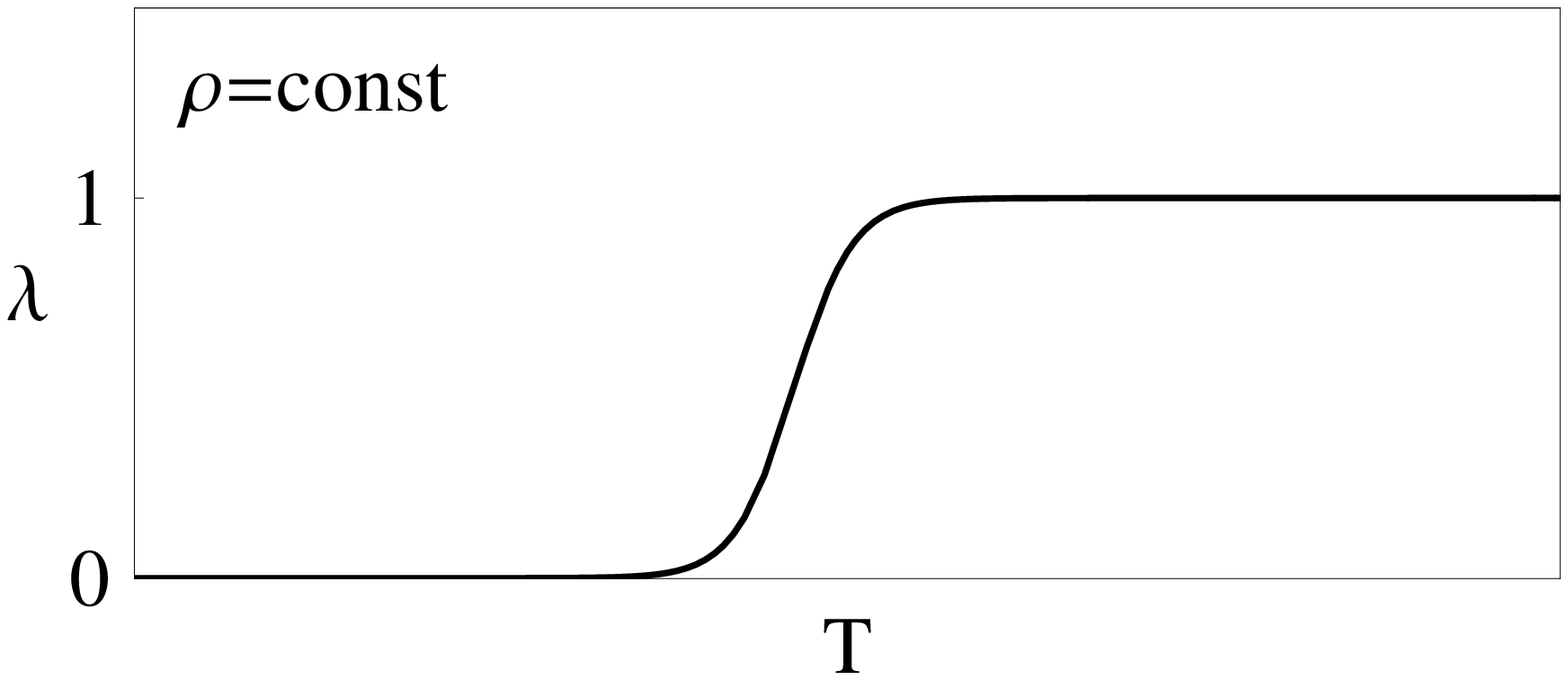}
 \caption{Schematics view on the mixed phase boundaries in the $\{T,\mu\}$
 (on left) and $\{T,\rho\}$ (on right) representations. Boundaries for
 a system conservation of a single charge (top), two charges (middle) as
 well as the volume fraction (bottom) are presented~\cite{KS3T06}.}
  \label{3Dbnd}
\end{figure}

Along with pressure, the temperature as well as the baryonic and
electric chemical potentials do not remain constant within the
mixed phase. This behavior is responsible for disappearance of the
entropy discontinuity  at the given $p$ and  $x$ as pictured in
the right panel of Fig.\ref{S_T}. This fact gave grounds for
author~\cite{M97} to claim that in the iso-asymmetric matter the
first-order phase transition is smoothed and becomes  the
second-order phase transition.  One should note that in spite of
such behavior of the entropy, the first derivative of the
thermodynamic potential with respect to temperature at the
constant chemical potentials (i.e. locally in the $\mu_i$ space)
suffers a jump (see Fig.\ref{Smu_T}), what is evidence of the
first-order phase transition in iso-asymmetric matter. At the same
time if the $\{p,\rho_B, x\}$ representation of entropy is used,
the result of~\cite{M97} is well reproduced (see the left panel of
Fig.\ref{S_T}). This mirrors a general property inherent to
systems with the first-order phase transition: In the mixed phase
the first derivatives of the thermodynamic potential have
discontinuity in the $\{T, \mu_{1},..., \mu_{n}\}$ representation
but they are continuous in the $\{T, \rho_{1},..., \rho_{n}\}$
one. That is the direct consequence of the Gibbs phase-equilibrium
conditions~\cite{LL69} from which follows that in the first
representation the every point of the $n$-dimensional mixed phase
surface  maps unambiguously on the $\{T, \mu_{1},..., \mu_{n}\}$
space. Evidently, it is not the case in the $\{T, \rho_{1},...,
\rho_{n}\}$ representation~\cite{KS3T06} as illustrated in
Fig.\ref{3Dbnd}. Due to that the volume fraction of the second
phase $\lambda\equiv \frac{V_{II}}{V}$ jumps when the system
enters in the mixed phase, if the first representation is used,
while it does not in the second representation. Any extensive
thermodynamic quantity $A$ will follow this behavior of $\lambda$
since $A=\lambda A_{II}+(1-\lambda) A_{I}$.

\begin{figure}[h]
 \includegraphics[width=12.cm]{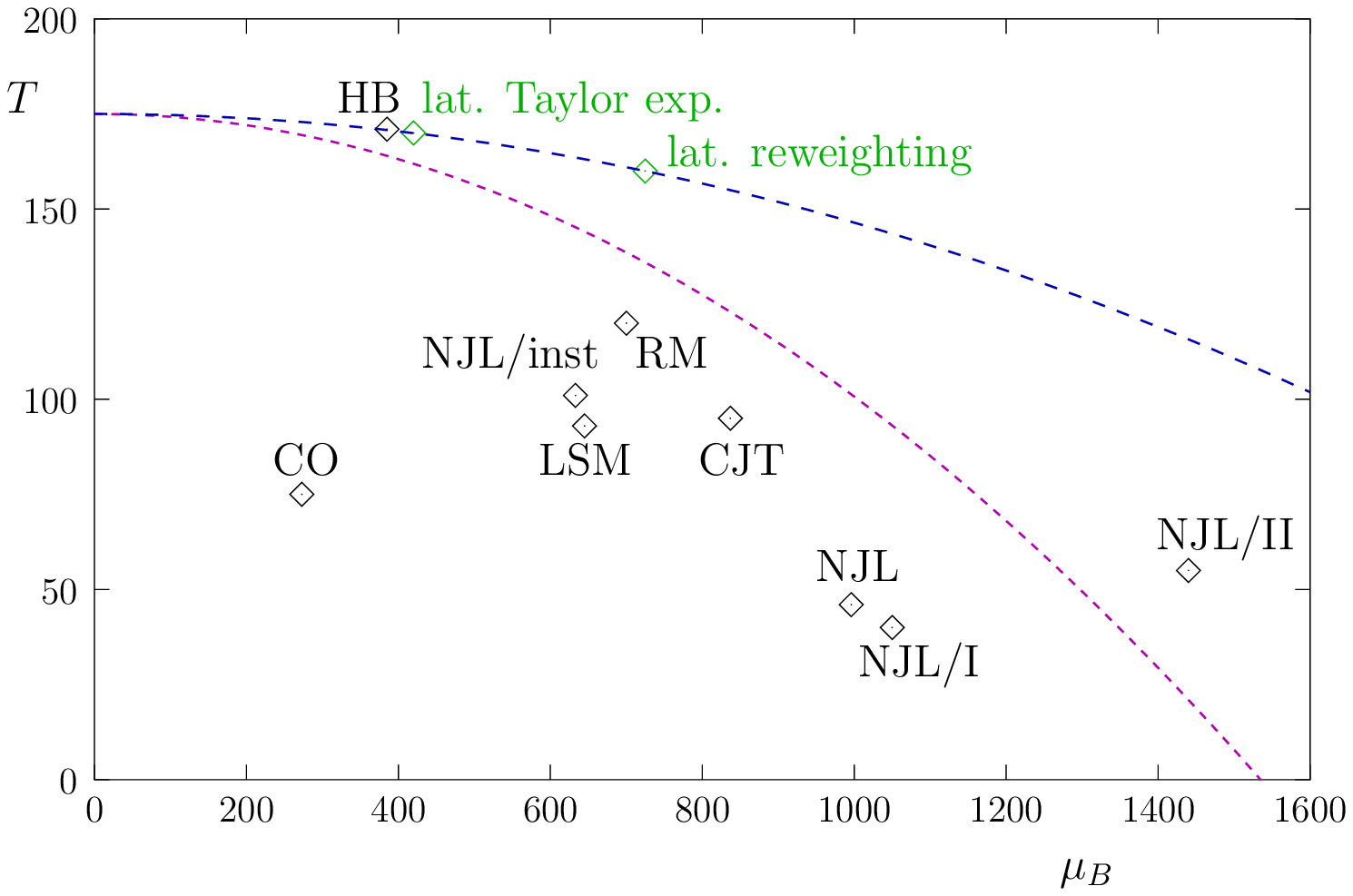}
  \caption{Theoretical (model and lattice)  predictions for the location
  of the critical end-point. Two lines are obtained by lattice Taylor
  expansion where the lower curve corresponds to smaller quark
   masses (compare with Fig.\ref{fig2-3j}). Points are calculated
   in different models specified in~\cite{St04}.
   Errors/uncertainities are not shown.
   }
  \label{cr_point}
\end{figure}


 As was noted above, the presence of two conserving charges in
the $\{T, \mu_B, \mu_{Q}\}$ representation changes dimensionality
of the two-phase coexistence surface for the first-order phase
transition. Namely, the one-dimensional line for the case of a
single conserving charge transforms into a two-dimensional (or
$n$-dimensional) (hyper)surface if two (or $n$) charges are
conserved~\cite{LL69}, see Fig.\ref{3Dbnd}. Therefore, the
manifold of critical points, that by definition is the boundary of
the mixed phase where the system suffers the second order phase
transition, also changes sufficiently: From two isolated points
defining the limited line of the mixed phase it transforms into a
one-dimensional curve ($(n-1)$-dimensional (hyper)surface)
being a boundary of two-dimensional ($n$-dimensional)
coexistence (hyper)surface~\cite{KS3T06}. The different topologies of the
mixed phase states may result in different important consequences.
In particular, in the discussed now new projects of
FAIR~\cite{GSI300} and RHIC~\cite{RHIC06} aimed to search for
manifestation of the critical end-point, being the point where the
first-order phase transition of the QCD matter ends, it may turn
out that it is not a point but hypersurface those dimensionality
is defined by a number and properties of conserving
charges~\cite{KS3T06}. It is noteworthy that the location of the
critical end-point in the $T-\mu_B$ plane was estimated in both
various model and lattice QCD calculations, however, without
taking into account isospin asymmetry degree of freedom. But even
in this case the predictions vary wildly as demonstrated in
Fig.\ref{cr_point} taken from \cite{St04}.

\begin{figure}[h]
 \includegraphics[width=12.cm]{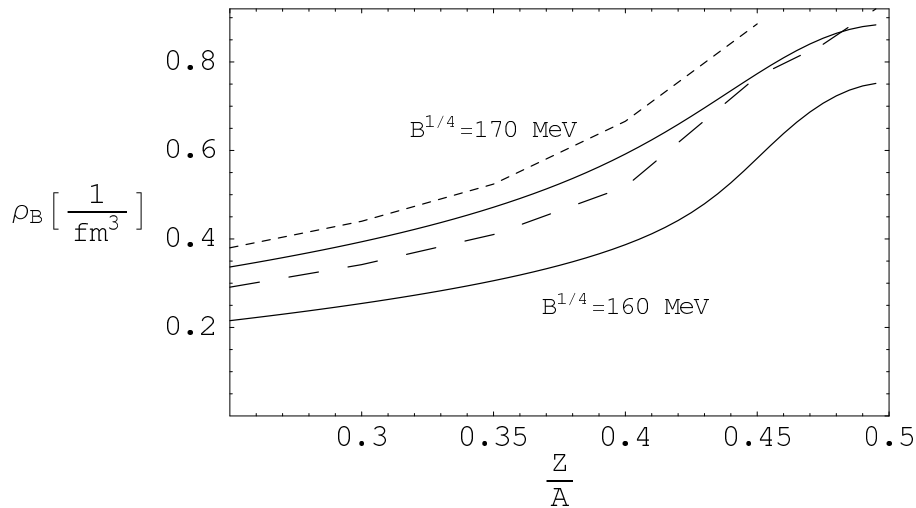}
  \caption{Transition density as a function of $Z/A$ at $T=50$
  MeV~\cite{DiTDGGL06}. The solid lines are obtained neglecting
  finite-size effects. The results plotted as long and short-dashed
  lines take into account these effects and
  correspond to $B^{1/4}=160$ MeV and $B^{1/4}=170$ MeV, respectively.}
  \label{fig13}
\end{figure}

The outstanding problems involved in understanding the
hadron-quark matter phase transition, aside from the description
of these phases themselves, are related to the geometric structure
of the mixed phase and its evolution with varying the relative
fraction $\lambda$ of phases. If one of the conserved charges is
the electric charge, a geometric structure in the mixed phase is
expected. The nucleation mechanism for cluster formation (say,
quark drops in the hadron environment) is dominant in the
metastable region of the first-order phase transition and
formation of the mixed phase influenced by finite size effects due
to nonvanishing surface tension at the interface between the
hadronic and quark matter and the Coulomb energy of the formed
drops. The geometric structure of the mixed phase was studied in
some detail for neutron star matter (see the
review-article~\cite{Nstar}) but for nuclear matter uncertainties
are rather large. In Fig.\ref{fig13}, an estimate of the
finite-size effects is presented for the nuclear
case~\cite{DiTDGGL06}.

\begin{figure}[h]
 \includegraphics[width=9.cm,angle=90]{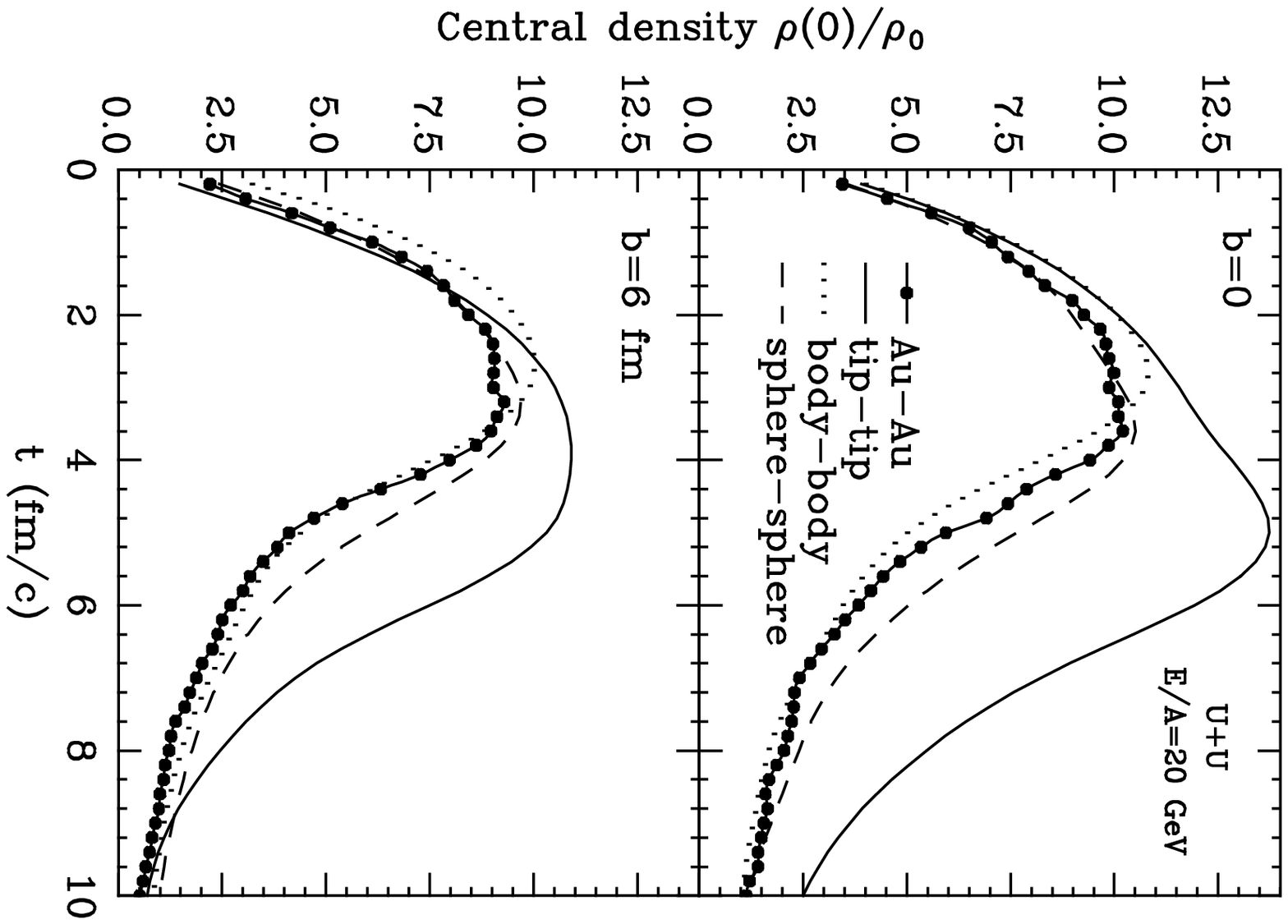}
 \includegraphics[width=9.cm,angle=90]{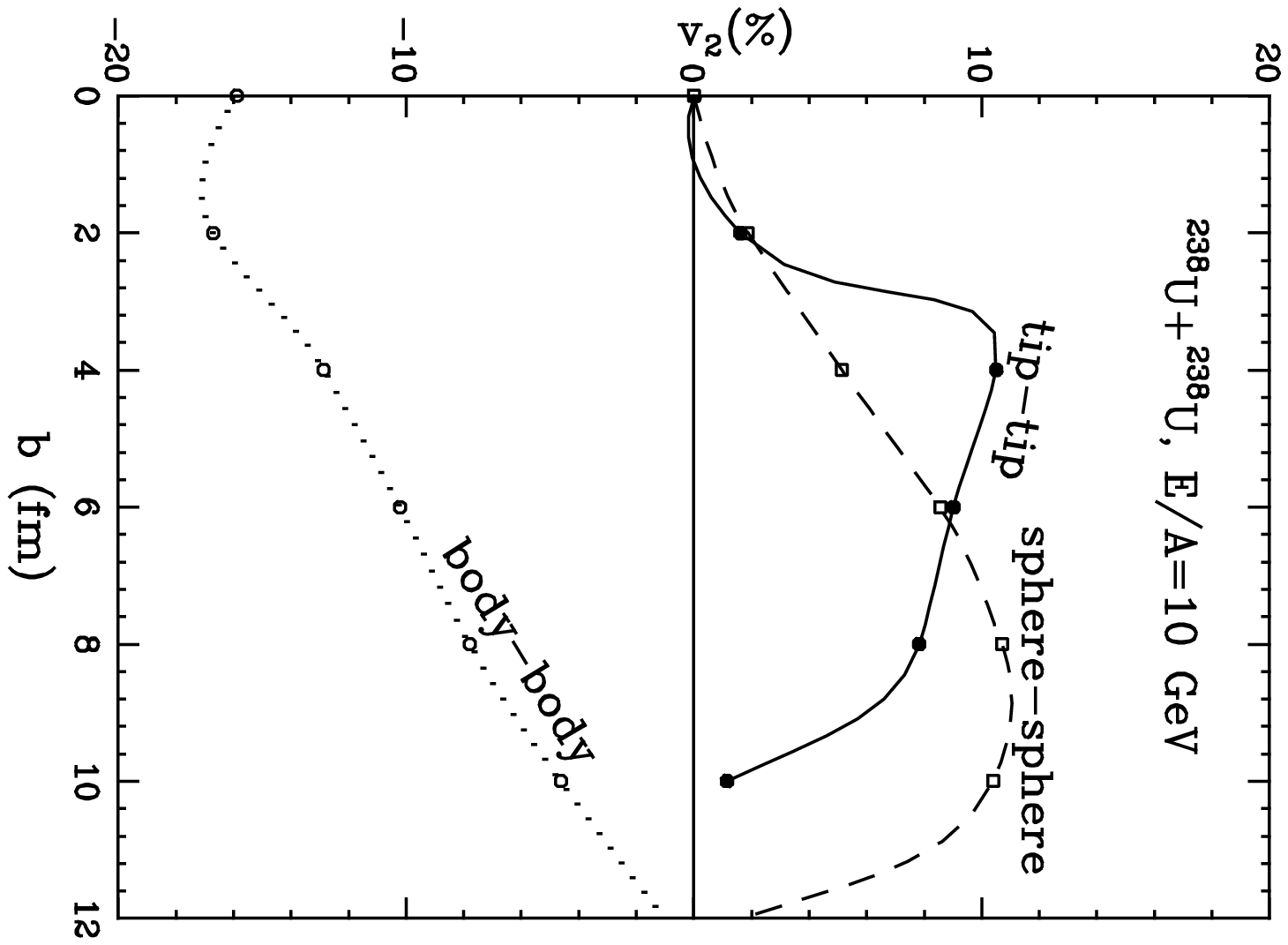}
 \caption{Evolution of central baryon density for Au+Au and U+U
 collisions at 20A GeV (left panel) and  impact parameter
 dependence of nucleon elliptic flow at 10A GeV (right panel)
 for different orientations of deformed Uranium nuclei as well
  as for spherical ones~\cite{BAL00}.}
  \label{def}
\end{figure}

As is clearly seen, the value of the density at which the mixed
phase can be reached becomes larger, if the finite size effects
are taken into consideration. This increase  due to finite-size
effects is larger for smaller values of the bag constant $B$
(compare with results in Fig.\ref{bnd}). Nevertheless, a strong
dependence of the transition energy on $Z/A$ survives even in this
case~\cite{DiTDGGL06}.

Besides the isospin there is yet another nuclear parameter which
may influence evolution of a colliding system. It is the nuclear
shape. It was noted earlier~\cite{BAL00,deform} that deformation
and orientation affect compression, elliptic flow and particle
production for collisions of Uranium nuclei. As  seen from
Fig.\ref{def}, the compression in the tip-tip U+U collisions is
about 30$\%$ higher and the region with $\rho > 5\rho_0$ lasts
approximately 40$\%$ longer than in the body-body collisions or
spherical U+U collisions. Moreover, the nucleon elliptic flow has
some unique features which in principle may allow one to
disentangle these two orientations \footnote{We are thankful to
N.~Xu for drawing our attention to this issue.}. Such situation is
valid for the energy range 1-20A GeV~\cite{BAL00} but
uncertainties are still large~\cite{deform}.

\section{On signals and precursors}

Similarly to the general situation with searching for a
quark-gluon plasma in ultrarelativistic collisions, there is no
single crucial experiment which unambiguously solves the problem.
It is quite evident that direct information on the mixed phase may
be obtained only by means of weakly interacting photon and lepton
probes.
 Sensitivity of global hadron observables to possible phase transitions
 is expected to be weak, but it might not  be the case for more delicate
characteristics. In any case, due to the proximity of the phase
diagram region under discussion to the confinement transition and
chiral symmetry restoration, some precursory phenomena cannot be
excluded at the bombarding energy below 10A GeV, which opens a new
perspective for physical investigations at the Dubna Nuclotron.

 Properties of hadrons  are expected to change in hot and/or
dense baryon matter~\cite{TK94,BR96}.  This change concerns
hadronic masses and widths, first of all those for the
$\sigma$-meson as the  chiral partner of pions which characterizes
a degree of chiral symmetry violation and can serve as a "signal"
of its restoration as well as the mixed phase formation. Rare
decays in matter of vector mesons (particularly $\rho $ and
$\omega$) are also very attractive.

\begin{figure}[h]
 \includegraphics[width=7.5cm,angle=-90]{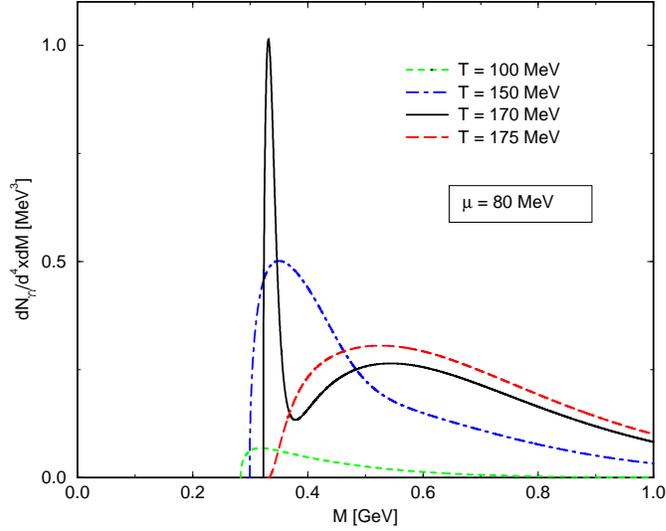}
  \caption{Invariant mass spectra of $2\gamma$ at $\mu_B=80$ MeV
  and different temperatures~\cite{VKBRS98}. }
  \label{fig6j}
\end{figure}

Nevertheless, there are theoretical proposals to probe chiral
symmetry restoration in the vicinity of the phase transition
boundary. In particular, it was shown~\cite{CH98,VKBRS98} that a
two-photon decay of the $\sigma$-meson formed as an intermediate
state in $\pi\pi$ scattering may be a very attractive signal. As
depicted in Fig.\ref{fig6j}, at temperature in the vicinity of
 the phase transition, when $m_\sigma \sim 2m_\pi$, there is an anomalous peak
in invariant mass spectra of $\gamma$ pairs which may serve as a
signal of the phase transition and formation of a mixed phase.
Certainly, there is a huge combinatorial background due to $\pi^0
\to \gamma\gamma$ decays, but the Nuclotron energy is expected to
have some advantage against higher energy accelerators because the
contribution of gamma's from deconfined quarks-gluons will be
negligible.

\begin{figure}
\includegraphics[width=6.5cm,angle=-90]{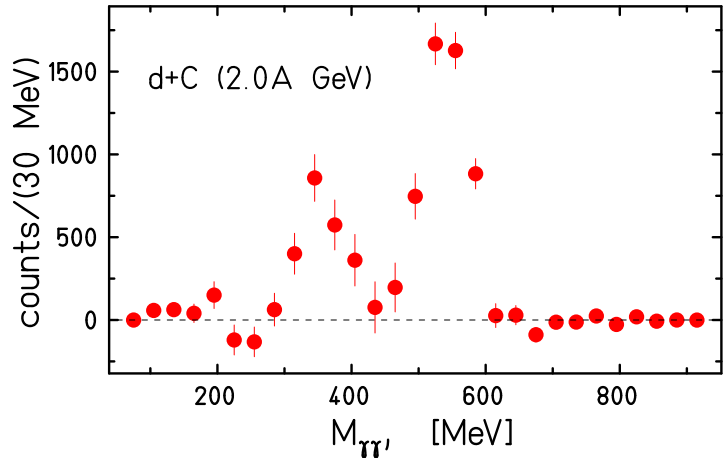}
  \caption{Invariant mass distribution of  pairs of $\gamma$ quanta with
  the energy $E_\gamma >100$ MeV after subtraction
  of the combinatorial background in the $d+C\to \gamma + \gamma +X$
  reaction at the bombarding energy 2A GeV~\cite{ASS06}. }
  \label{figdC}
\end{figure}

To measure this signal, a photon spectrometer is needed. The
installation of this kind, two-arm $\gamma$-spectrometer PHOTON-2,
is available at JINR. The ability of this installation in
discriminating  two correlated photons is demonstrated in
Fig.\ref{figdC}.

The presented invariant mass spectrum of $\gamma\gamma$
correlations is based on $1.5\cdot 10^{6}$ selected
dC-interactions. A peak corresponding to about 5000 $\eta$-meson
decays is clearly seen at $M_\eta =540.5\pm 2.1$ MeV with the
resolution $\sigma=33.4\pm6.0$~\cite{ASS06}. The measured $\eta$
mass is consistent with the table value. A huge peak from the
$\pi^0\to \gamma\gamma$ decay is strongly suppressed here and is
not visible in the figure due to the selection criteria of events
to increase the signal/background ratio. It is of interest that
some additional structure with a maximum in the range of 340-460
MeV is seen which was not observed in analogous experiments with
lower statistics~\cite{ASS06}. The nature of this resonance
structure is under discussion now.

Fluctuations in particle number are inherent in the first-order
phase transition and formation of a metastable mixed phase.
However, to a certain extent, peculiarities of a phase transition
will be washed out by transition dynamics and subsequent hadron
interactions which drive the system closer to equilibrium, which
means the loss of information on the system prehistory.
Nevertheless, it was understood that
  the  study of fluctuations in relativistic
strongly interacting matter may help with solving the problems
mentioned above. Experimental data on event-by-event fluctuations
(e.g., fluctuations in particle multiplicity, electric, baryon and
strangeness charges) in nuclear collisions give a unique
possibility to test recent ideas on the origin of fluctuations in
relativistic interacting systems~\cite{Fluct1,Fluct2}. Among them
the suppression of event-by-event fluctuations of electric charge
 was predicted \cite{Fluct1} as a consequence of deconfinement.
Theoretical estimates of the magnitude of the charge fluctuations
indicate that they are much smaller in a quark-gluon plasma than
in a hadron gas. Thus, naively, a decrease in the fluctuations is
expected when the collision energy crosses the threshold for the
deconfinement phase transition. However, this prediction is
derived under assumptions that  initial fluctuations survive
through hadronization and that their relaxation times in hadronic
matter are significantly longer than in the hadronic stage of the
collision \cite{Fluct1,Zar02}.

More realistic estimates have been done recently for the hadron
resonance-gas model within a statistical approach~\cite{BGHKZ06}.
The equilibrium statistical model was successfully used to
describe the data on hadron multiplicities in relativistic
heavy-ion collisions~\cite{BRS03}.  The system was treated in
Grand Canonical (GCE) and Canonical (CE) Ensembles with taking
into account the conservation of baryon, electric and strangeness
charges. The decay of resonances was included in fluctuations
which were calculated for nuclear states to be realized along the
experimental freeze-out curve. The whole available energy range,
from SIS to RHIC energies, was covered in this
study~\cite{BGHKZ06}.

\begin{figure}[thb]
\includegraphics[width=7.9cm,height=7.cm]{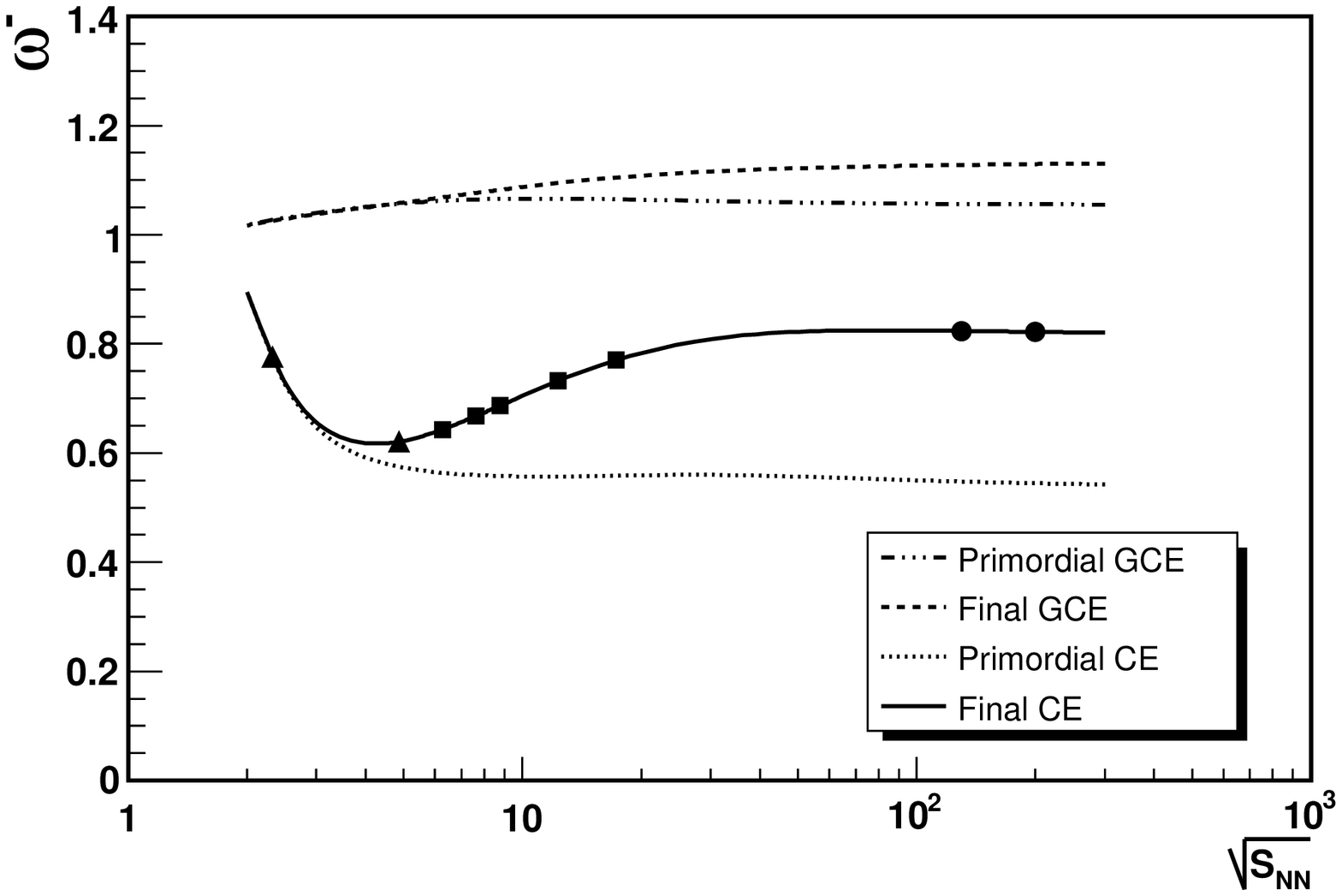}
\includegraphics[width=7.9cm,height=7.cm]{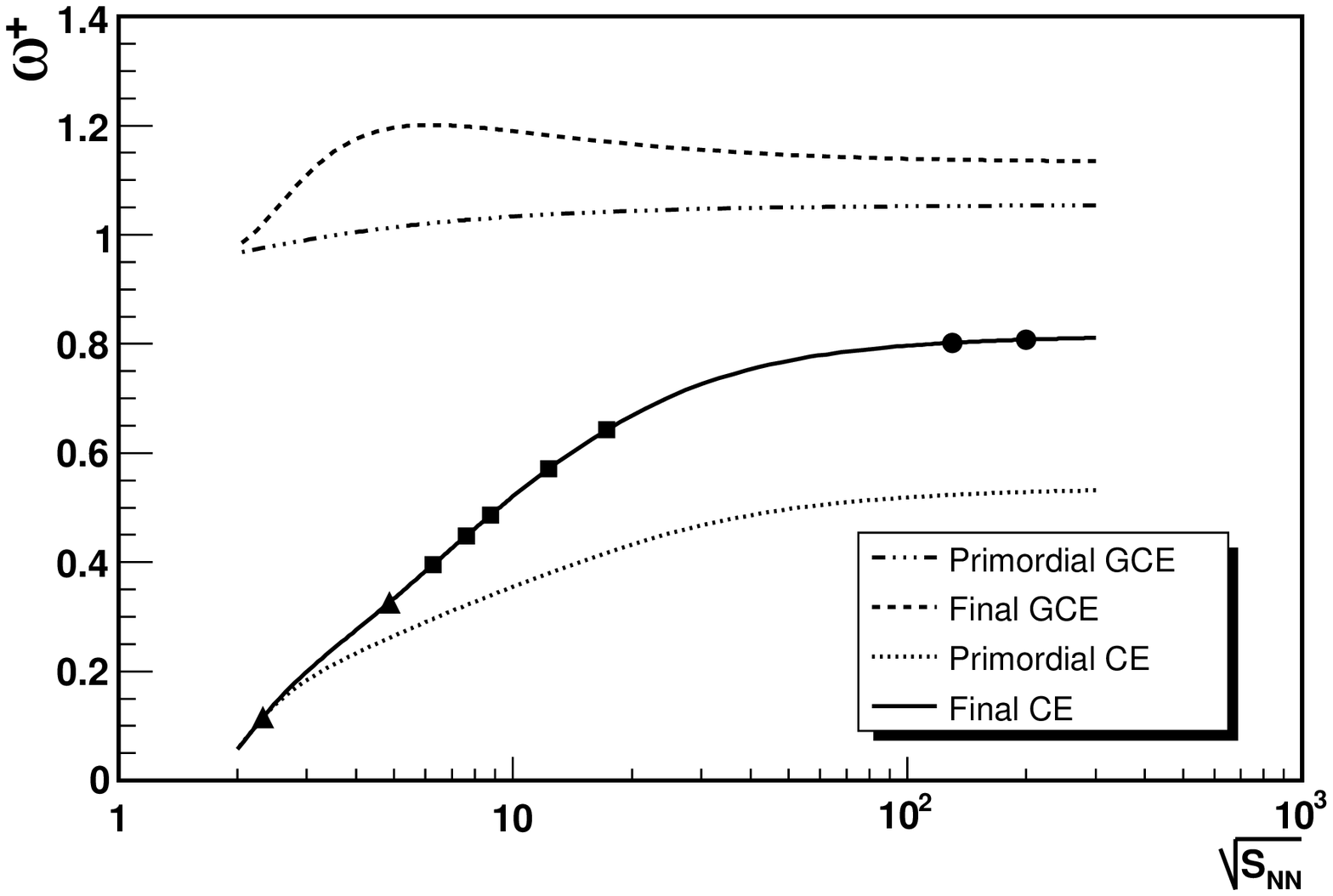}
   \caption{The colliding energy dependence of the scaled
   variances for negatively $\omega^-$ (left panel) and positively
   $\omega^+$ (right panel) charged particles calculated along the
   chemical freeze-out line for central Pb+Pb (Au+Au)
   collisions~\cite{BGHKZ06}. Different lines present primordial
   and final (i.e. including resonance decays)  GCE and CE results.
   Solid points show the actual energy
   values of AGS (triangles), SPS (squires) and RHIC (dots) .
    }
   \label{fig-fluct}
\end{figure}

We exemplify these predictions by the results in
Fig.\ref{fig-fluct} for the scaled variance of negatively and
positively charged hadrons,
\begin{eqnarray}
\omega^-=  <(\Delta N^-)^2>/ <N^-> \ \ \mbox{and} \ \ \omega^+=
<(\Delta N^+)^2>/ <N^+>~,  \nonumber
\end{eqnarray}
 where $N^\pm$ is the number of particles in the given event
 and $<...>$ denote averaging over all events.

As is seen from the comparison of the GCE and CE results in
Fig.\ref{fig-fluct}, the exact charge conservation is very
important in the Nuclotron energy. While the energy $E_{lab} \sim
10$A GeV is approached, the resonance decay is getting sizable.
There is essentially different behavior of the scaled variances
 for negative, $\omega^-$, and positive, $\omega^+$, charges in the
 Nuclotron energy range $E_{lab}\lsim 10$A GeV~\cite{BGHKZ06}.

At a first glimpse, these statistical results may serve as a
reference point to look for peculiarities related to a possible
formation of the mixed phase. However, it is not the case due to
 at least two reasons. First, in addition to the considered statistical
fluctuations, heavy-ion collisions generate dynamical
fluctuations. Secondly, the above presented results correspond to
an ideal situation when all final hadrons are accepted by the
detector. Indeed, as was shown in~\cite{KGBS06} with using a
transport approach to heavy-ion collisions, the fluctuations in
the initial energy deposed in the statistical system yield
dynamical fluctuations of all macroscopic parameters. In its turn,
these fluctuations are dominated by a geometric variation of the
impact parameter. However, even for the fixed impact parameter a
number of participants fluctuates from event to event. The
centrality of the selected events is commonly controlled by a
number of measured projectile spectators. It was noted
in~\cite{KGBS06} that this procedure introduces some asymmetry
between fluctuations in target and projectile, thereby the scaled
variances $\omega^-$ and $\omega^+$ behave differently in the
target and projectile rapidity regions. Thus, the above noted
first and second reasons are turned out to be closely related. A
first attempt to compare these results with the NA49 data at
$E_{lab}=158$A GeV shows a remarkable disagreement but
unfortunately it is still impossible to make any conclusions from
this fact due to large uncertainties in these experimental
data~\cite{KGBS06}. Note that the considered approaches take into
account no phase transition.  To search for signals of the mixed
phase in fluctuations, the observable effect should be tested
against its variations in bombarding energy, centrality selection
and isospin asymmetry of colliding nuclei.

 One should stress that measurements of fluctuations require tracking
 detectors of  large acceptance, good particle identification  and a
 precise control of collision
 centrality on event-by-event basis. Previous experiments suffered either
 small phase-space coverage or limited tracking and particle identification,
 therefore  new measurements at the Nuclotron energy are of
particular importance. From the experimental point of view the
Nuclotron energy range seems to be ideal for these measurements.
This is because moderate particle multiplicity and their
relatively broad angular distribution simplify an efficient
detection of all produced charged particles.

A particular property of the mixed phase is the so-called
distillation effect: While the total charge is conserved, its
distribution between two phases is different. The structure of the
mixed phase is especially rich and complicated if the conservation
of the baryon number, electric charge and strangeness are taken
into consideration simultaneously. The distillation effect is
expected to result in certain observable effects as discussed
in~\cite{BSGDiT05,TNFNR04}.

One should note that the use of neutron-rich isotopes or,
generally, radioactive beams has recently become a central theme
in nuclear and astro-physics researches. At present, there are
several facilities devoted to nuclear physics studies using {\it
low-energy} beams of radioactive species including the Spiral at
Ganil (Caen, France), ACCULINNA and DRIBs (Dubna, Russia),
Cyclotron Research Center (Louvain-la-Neuve, Belgium), TRIUMF
(Vancouver, Canada), Holifield Radioactive Ion Beam Facility (Oak
Ridge, USA). These capabilities for radioactive beams will
eventually allow for detailed studies of the structure of nuclei
on the path of astrophysical $r$-process and provide for
fundamental nuclear structure studies of very neutron-rich nuclei
but with small atomic numbers. Heavy-ion radioactive beams are
also of great interest for both nuclear structure studies and some
astro-physics problems, in particular, clarification of the role
of new $\nu p$-process in nucleosynthesis of nuclei with $A\sim$
100~\cite{FML06}. To our knowledge, there is no project to get
accelerated heavy neutron-rich nuclei. Unfortunately, heavy
nuclear isotopes which can be obtained in a reasonable amount
cannot really be extended beyond the region of the isospin
asymmetry parameter $0.39\lsim Z/A\lsim 0.41$ reachable with
stable nuclei.

\section{Conclusions}

A study of the phase diagram of strongly interacting QCD matter in
the domain populated by heavy-ion collisions with the bombarding
energy $\lsim 10$A GeV and a search for manifestation of the mixed
phase formation seem to be a very attractive task. The use of the
isospin asymmetry as an additional conserving parameter to
characterize the created hot and dense system draws new interest
in this problem. Unfortunately, the available theoretical
predictions are strongly model dependent giving rather dispersive
results. There are no lattice QCD predictions for this highly
nonpertubative region. So much theoretical work should be done and
only future experiments may disentangle these models.

 A JINR Nuclotron possibility of accelerating heavy ions to the
project energy of 5A GeV and increasing it up to 10A GeV can be
realized in about two-three years. This will enable us to effort a
unique opportunity for scanning heavy-ion interactions in energy,
centrality and isospin asymmetry of the system to search for the
mixed phase of QCD matter. For the latter point it seems to be
optimal to have the gold and uranium beams in order to scan in
isospin asymmetry in both central and semi-central collisions at
not so high temperatures favorable for observation of the phase
boundary reduction. The use of strongly deformed U nuclei is quite
promising to probe the orientation effect in heavy ion collisions.
All this gives a chance to address experimentally many recent
problems within the next several years before the FAIR GSI
accelerator comes into operation. Being supplemented by scanning
in the isospin asymmetry parameter, as discussed in the present
paper,  the proposed research program at the
Nuclotron~\cite{SSSTZ06} may be considered also as a pilot study
preparing for subsequent detailed investigations at
SIS-100/300~\cite{GSI300} and as an integral part of the world
scientific cooperation to study the energy dependence of hadron
production properties in nuclear collisions.

\vspace*{3mm}
 {\bf Acknowledgements}
We greatly appreciate many useful and valuable discussions on
physics, detectors  and accelerator technique with
Kh.Y.~Abraamyan,  J.G.~Brankov, Yu.P.~Gangrsky, M.~Gazdzicki,
M.I.~Gorenstein, G.G.~Gulbikyan,  H.~Gutbrod, T.~Hollman,
M.G.~Itkis, Yu.B.~Ivanov, A.S.~Khvorostukhin, A.D.~Kovalenko,
R.~Lednicky, A.I.~Malakhov, I.N.~Meshkov, Yu.E.~Penionzhkevich,
V.B.~Priezzhev, V.N.~Russkikh, Yu.M.~Sinyukov, V.V.~Skokov,
M.K.~Suleymanov, G.M.~Ter-Akopyan, D.N.~Voskresensky, N.~Xu, and
G.M.~Zinovjev.
 We would like also to express our special thanks to V.G.~Kadyshevsky,
 V.A.~Matveev, and A.N.~Tavkhelidze for their interest
 to this paper.

This work was supported in part by RFBR Grant N 05-02-17695 and by
the special program of the Ministry of Education and Science of
the Russian Federation (grant RNP.2.1.1.5409).


\begin{thebibliography}{99}

\bibitem{GSI300}Proposal for an International Accelerator Facility
for Research with Heavy Ions and Antiprotons,
http://www.gsi.de/documents/DOC-2004-Mar-196-2.pdf~.
%
%
\bibitem{RHIC06}G.S.F. Stephans,
arXiv:nucl-ex/0607030; A.Cho, Science {\bf 312}, 190 (2006).
%
\bibitem{SSSTZ06} A.N.~Sissakian, A.S.~Sorin, M.K.~Suleymanov, V.D.~Toneev,
and G.M.~Zinovjev, arXiv:nucl-ex/0601034; Proceedings of the 8th
International Workshop "Relativistic Nuclear Physics: From
hundreds MeV to TeV",~(Dubna, May 23 - 28, 2005), Dubna, 2006,
p.306 [arXiv:nucl-ex/0511018].
%
\bibitem{IRT05} Y.B.~Ivanov, V.N.~Russkikh and V.D.~Toneev,
Phys. Rev. {\bf C 73}, 044904 (2006) [arXiv:nucl-th/0503088].
%
\bibitem{KSTR06} A.S.~Khvorostukhin, V.V.~Skokov, V.D.~Toneev and K.~Redlich,
arXiv:nucl-th/0605069.
%
\bibitem{Fodor01}
Z. Fodor and S.D. Katz, JHEP {\bf 203}, 14 (2002)
[arXiv:hep-lat/0106002]; JHEP {\bf 404}, 50  (2004)
[arXiv:hep-lat/0402006].
%
\bibitem{ZS} E.V.~Shuryak and I.~Zahed, arXiv:hep-ph/0307267.
%
\bibitem{V04}D.N.~Voskresensky,
Nucl. Phys. A {\bf 744},  378 (2004) [arXiv:hep-ph/0402020];
 G.E.~Brown, Ch,-H.~Lee, and M.~Rho,
arXiv:hep-ph/0402207.
%
\bibitem{I05} Yu.~Ivanov, Multi-fluid
hydrodynamics, Talk at the CBM Collaboration Meeting {\it "FAIR,
The physics of compressed baryonic matter"}, December 15-16, 2005,
GSI, Darmstadt,
http://www.gsi.de/documents/DOC-2005-Dec-87-112-1.pdf;
V.N.~Russkikh, privite communication.
%
\bibitem{LL69} L.D.~Landau and E.M.~Lifshitz, {\it Statistical
Physics} (Addison-Wesley, Reading, MA, 1969), Chap. VIII, IX.
%
\bibitem{G92} N.K.~Glendenning, Phys. Rev. {\bf D46} 1274 (1992).
%
\bibitem{M97} H.M\"uller,
Nucl. Phys. {\bf A 618}, 349 (1997) [arXiv:nucl-th/9701035].
%
\bibitem{KTV06} A.S~Khvorostukhin, V.D.~Toneev and
D.N.~Voskresenski (to be published).
%
\bibitem{KV05} E.E.~Kolomeitsev and D.N.~Voskresenski, Nucl. Phys. {\bf A759}, 373 (2005)
[nucl-th/0410063].
%
\bibitem{DiTDGGL06} M. Di~Toro, A.~Drago, T.~Gaitaos, V.~Greco and
A.~Lavagno,
arXiv:nucl-th/0602052.
%
\bibitem{GM91}N.K.~Glendenning and S.A.~Moszkowski,
Phys. Rev. Lett. {\bf 67}, 2414 (1991).
%
\bibitem{LGBCDiT02}B.~Liu, V.~Creco, V.~Baran, M.~Colonna and
M.~Di~Toro,  Phys. Rev. {\bf C65}, 045201 (2002).
%
\bibitem{W84}E.~Witten, Phys. Rev. {\bf D30}, 272 (1984).
%
\bibitem{KS3T06} A.S~Khvorostukhin, A.N.~Sissakian, V.V.~Skokov,
A.S.~Sorin, and V.D.~Toneev (to be published).
%
\bibitem{St04}M.~Stephanov,
Int. J. Mod. Phys. {\bf A20}, 4387 (2005)[arXiv:hep-ph/0402115].
%
\bibitem{Nstar} T.~Maruyama, T.~Tatsumi, T.~Endo and S.~Chiba,
arXiv:nucl-th/0605075.
%
\bibitem{BAL00} Bao-An Li,
Phys.Rev. {\bf C61}, 021903 (2000) [arXiv:nucl-th/9910030].
%
\bibitem{deform}E.V. Shuryak,
 Phys.Rev. {\bf C61}, 034905 (2000) [arXiv:nucl-th/9906062];
A.J. Kuhlman and U.W. Heinz,
Phys.Rev. {\bf C72}, 037901 (2005) [arXiv:nucl-th/0506088]; C.
Nepali, G. Fai and D. Keane,
Phys. Rev. {\bf 73}, 034911 (2006) [arXiv:hep-ph/0601030].


%
\bibitem{TK94}T.~Hatsuda and T.~Kunihiro, Phys. Rep. {\bf 247}, 221(1994).
%
\bibitem{BR96} G.E.~Brown and M.~Rho, Phys. Rep. {\bf 269}, 333 (1996).
%
\bibitem{CH98}S.~Chiku and T.~Hatsuda, Phys. Rev. {\bf D58}, 076001 (1998)
[arXiv:hep-ph/9809215]; T.~Hatsuda, T.~Kunihiro and H.~Shimizu,
Phys. Rev. Lett. {\bf 82}, 2840 (1999).
%
\bibitem{VKBRS98}  M.K.~Volkov, E.A.~Kuraev, D.~Blaschke, G.~R\"opke and
S.M.~Schmidt, Phys. Lett. {\bf B424}, 235 (1998)
[arXiv:hep-ph/9706350].
%
\bibitem{ASS06}Kh.U.~Abraamyan, A.N.~Sissakian and A.S.~Sorin,
 arXiv:nucl-ex/0607027.

%
\bibitem{Fluct1} S.~Jeon and V.~Koch, Phys. Rev. Lett. {\bf 85}, 2076 (2000)
[arXiv:hep-ph/0003168]; M.~Asakawa, U.W.~Heinz and B.~Muller,
Phys. Rev. Lett. {\bf 85}, 2072 (2000) [arXiv:hep-ph/0003169];
E.V.~Shuryak and M.A. ~Stephanov, Phys. Rev. C {\bf 63}, 064903
(2001) [arXiv:hep-ph/0010100].
%
\bibitem{Fluct2} V.V.~Begun, M.~Gazdzicki, M.I.~Gorenstein, and O.S.~Zozulya,
Phys. Rev. {\bf C70}, 034901 (2004) [arXiv:nucl-th/0404056];
V.V.~Begun, M.I.~Gorenstein, A.P.~Kostyk and O.S.~Zozulya,
Phys. Rev. {\bf C 71}, 054904 (2005) [arXiv:nucl-th/0410044];
A.~Keranen, F.~Becattini V.V.~Begun, M.I.~Gorenstein, and
O.S.~Zozulya,
J. Phys. {\bf G 31}, S1095 (2005) [arXiv:nucl-th/0411116].
%
\bibitem{Zar02} J.~Zaranek, Phys. Rev. C {\bf 66}, 024905 (2002)
[arXiv:hep-ph/0111228].
%
\bibitem{BGHKZ06}
V.V.~Begun, M.I.~Gorenstein, M.~Hauer, V.P.~Konchakovski and
O.S.~Zozulya,
arXiv:nucl-th/0606036.
%
\bibitem{BRS03}P. Braun-Munzinger, K. Redlich and J. Stachel,
in {\em Quark Gluon Plasma 3} eds. R.C.~Hwa  and X.N.~Wang, World
Scientific, Singapore, p.491 [nucl-th/0304013];
 A. Andronic, P. Braun-Munzinger and J. Stachel, arXiv:nucl-th/0511071.
%
\bibitem{KGBS06} V.P.~Konchakovski, S.~Haussler, M.I.~Gorenstein,
E.L.~Bratkovsrkaya, M.~Bleicher, and H.~St\"ocker,
Phys. Rev. {\bf C73}, 034902 (2006) [arXiv:nucl-th/0511083];
V.P.~Konchakovski, M.I.~Gorenstein,
arXiv:nucl-th/0606047.
%
\bibitem{BSGDiT05} V.~Baran, M.~Colonna, V.~Greco and M.~Di~Toro,
Phys. Rep. {\bf 410}, 335 (2005).
%
\bibitem{TNFNR04} V.D.~Toneev, E.G.~Nikonov,  B.~Friman, W.~N\"orenberg,
and K.~Redlich,
  Eur. Phys. J. {\bf C32}, 399 (2004) [arXiv:hep-ph/0308088].
%
\bibitem{FML06} C. Fr\"ohlich, G. Marti\'nez-Pinedo, M. Liebend\"orfer, F.-K.
Thielemann, E. Bravo, W.R. Hix, K. Langanke, and N.T. Zinner,
 Phys. Rev. Lett. {\bf 96}, 142502 (2006) [arXiv:astro-ph/0511376].


\end{thebibliography}
\end{document}